\documentclass[aps,prc,reprint,showpacs,groupedaddress,onecolumn]{revtex4-1}

\usepackage{amsmath}
\usepackage{amssymb}
\usepackage{amsfonts}
\usepackage{bm}          
\usepackage[dvips]{graphicx,color}

\def\beq{\begin{equation}}
\def\eeq{\end{equation}}
\def\beqa{\begin{eqnarray}}
\def\eeqa{\end{eqnarray}}
\def\bsigma{\mbox{\boldmath$\sigma$}}

\begin{document}

\title{Two-Nucleon Scattering without partial waves using 
a  momentum space Argonne V18 interaction \footnote{Dedicated to Professor 
Henryk Wita\l a at the Occasion of his 60th Birthday}}

\author{S. Veerasamy}
\email{veerasam@ohio.edu,sveerasamy2011@gmail.com}
\author{Ch. Elster}
\email{elster@ohio.edu}
\affiliation{
Institute of Nuclear and Particle Physics,  and
Department of Physics and Astronomy, Ohio University, Athens, OH
45701}

\author{W. N. Polyzou}
\email{polyzou@uiowa.edu}
\affiliation{
Department of Physics and Astronomy, The University of Iowa, Iowa City, IA
52242}

\date{\today}

\begin{abstract}

We test the operator form of the Fourier transform of the Argonne V18
potential by computing selected scattering observables and all
Wolfenstein parameters for a variety of energies. These are compared
to the GW-DAC database and to partial wave calculations.  We represent
the interaction and transition operators as expansions in a spin-momentum
basis.  In this representation the Lippmann-Schwinger equation becomes
a six channel integral equation in two variables.  Our calculations
use different numbers of spin-momentum basis elements to represent the on- and
off-shell transition operators.  This is because different  
numbers of independent spin-momentum basis elements are required to expand 
the on- and off-shell transition operators.
The choice of on and off-shell spin-momentum basis elements is made so that
the coefficients of the on-shell spin-momentum basis vectors are simply related
to the corresponding off-shell coefficients.
 
\end{abstract}

\vspace{10mm}

\maketitle


\section{Introduction}

Among the so-called `high precision' nucleon-nucleon (NN) potentials
the Argonne V18 potential~\cite{Wiringa:1994wb} is the only one given
in a basis of spin-isospin operators, $O_n$, which are multiplied by
scalar functions, $V_n(r)$, of the relative distance $r$.  These
features are advantageous for some applications, however, there are
classes of problems where a momentum-space treatment is preferred.
These applications include relativistic few-nucleon scattering and
electroweak probes of few-nucleon systems.  While realistic
momentum-space nucleon-nucleon potentials are available, they are
either given in terms of partial-wave expansions, e.g.  the CD-Bonn
interaction~\cite{Machleidt:2000ge} and the Nijmegen interaction~\cite{Stoks:1994wp}, 
or they are limited to low
energies, like the interactions derived in chiral effective field
theory~\cite{Epelbaum:2004fk,Entem:2003ft}.

The experience in standard three-nucleon calculations that is based
on a partial-wave projected momentum-space
basis~\cite{Gloeckle:1995jg} shows that while this standard treatment
is quite successful at lower energies, the numerical realization of
the scattering equations becomes more tedious with increasing
energy. For a system of three bosons interacting via scalar forces it
already has been demonstrated that it is relatively easy to directly
calculate three-body scattering observables in a relativistic Faddeev
scheme without using partial wave projections
~\cite{Lin:2008sy,Lin:2007kg}.  Thus, it is natural to strive for
solving the three-nucleon Faddeev equations in a similar fashion.

The first step in this direction is solving the two-nucleon problem
based on an operator expansion instead of a partial-wave projected
momentum-space basis.  One of the advantages of using operator
expansions over partial waves is that the transition matrix elements
are reasonably smooth.  At higher energies, an accurate representation
of these smooth, but forward-peaked, functions requires a large number
of partial waves.  A converged partial-wave expansion of the
transition matrix elements implies that all of the high-frequency
oscillations from the large-$l$ partial waves must largely cancel. In
addition, to achieve these cancellations the computation of
the transition matrix elements requires the accurate computation of
oscillatory integrals.  At low-energies partial waves are preferred
because they replace the two-variable Lippmann-Schwinger equation by a
finite set of uncoupled one-variable equations.  For sufficiently high
energies the simplicity of a direct vector treatment of the
Lippmann-Schwinger equation has advantages, even though the integral
equation has an additional integration variable.  In this paper we
discuss the direct vector solution of the Lippmann-Schwinger equation
using a recent momentum-space treatment of the Argonne V18
potential~\cite{veerasamy:2100}. There the potential was only tested
in the calculation of the deuteron bound state, which is not sensitive
to the charge symmetry breaking parts of the AV18 potential. Solving
for the scattering observables provides a complete test of the
momentum-space representation of the AV18 potential presented in
Refs.~\cite{veerasamy:2100,saravanan-thesis}.

There have been several approaches formulating nucleon-nucleon (NN)
scattering without employing a partial-wave decomposition. A helicity
formulation related to the total NN spin was proposed
in~\cite{Fachruddin:2000wv}. The spectator equation for relativistic
NN scattering has been successfully solved in \cite{Ramalho:2006jk},
also using a helicity formulation. 

A three-dimensional formulation based on an operator expansion
was proposed and carried out for a chiral next-to-next-leading order
NN force and a standard one-boson-exchange potential in
~\cite{Golak:2010wz}.  A basic foundation for the latter rests on the
fact that the most general form of the NN interaction can be
represented as a linear combination of six linearly-independent
spin-momentum operators with scalar coefficient functions.  This
representation determines the spin-structure of the NN bound and
scattering states.  The Wolfenstein decomposition of the NN scattering
amplitude into five linear operators is dictated by rotational, space
reflection, spin-exchange symmetry, and time-reversal
invariance~\cite{wolfenstein-ashkin}. A sixth independent operator
with these symmetries does not exist on shell.
Ref.~\cite{Golak:2010wz} uses six linearly independent operators that
satisfy all of the symmetry requirements, but these operators become
linearly dependent on shell so the scattering amplitude was computed
from the off-shell result by the required continuity of the transition
matrix elements.
 
The vector treatment of the spins requires the computation of an
analytic expression for each spin-basis element at each quadrature
point.  A symbolic reduction technique, developed in
~\cite{saravanan-thesis}, automates the computation of these
expressions.  Our calculations use five independent spin-operators to
expand the on-shell transition matrix element and an additional sixth
operator to expand the off-shell transition matrix elements.  Based on
calculations using different choices of the sixth operator, we find
that using a sixth operator that changes sign under time reversal to
expand the off-shell transition matrix elements leads to a numerically
more stable discretization of the Lippmann-Schwinger equation when
compared to calculations using a sixth operator that has the
symmetries of the potential.  An important observation is that while a
time-odd operator appears in the expansion of the potential, the
potential itself remains invariant with respect to time reversal
because the expansion coefficients also contain time-odd components.
The expansion coefficient of the 
time-odd operator that we consider vanishes on-shell.

In Sec.~II we discuss the operator basis that we use to
expand the NN potential and derive in Sec.~III the resulting
Lippmann-Schwinger equation and show how we explicitly obtain the
on-shell scattering amplitude.  We then discuss our numerical
procedure and the extraction of the Wolfenstein amplitudes in
Sec.~IV. In Sec.~V we discuss the selected NN observables at different 
energies, and conclude in Sec.~VI.

\section{Operator Expansions} 

Momentum-space scattering calculations are performed using the
Lippmann-Schwinger integral equation.  It is a singular integral
equation with a compact kernel.  This means that the kernel can
be uniformly approximated by a finite dimensional matrix and that the
equations can be solved to any desired accuracy by solving a finite
system of linear equations.

The Lippmann-Schwinger equation has the well-known form
\beq
T(z) = V + V (z-H_0)^{-1} T(z) .
\label{d.1}
\eeq
We represent the operators in this equation by matrix elements
in the total momentum $\mathbf{P}$ and single-particle momentum,
$\mathbf{p} = \frac{1}{2} (\mathbf{p}_p- \mathbf{p}_n) $,   
boosted to the system rest frame with a Galilean boost.  Matrix element of
the interaction and transition operator have the form 
\beq
\langle \mathbf{P}',\mathbf{p}' \vert V \vert \mathbf{P},\mathbf{p}
\rangle =
\delta (\mathbf{P}'-\mathbf{P}) \langle \mathbf{p}' \Vert V 
\Vert \mathbf{p} \rangle 
\label{d.2}
\eeq
and 
\beq
\langle \mathbf{P}',\mathbf{p}' \vert T(z) \vert \mathbf{P},\mathbf{p}
\rangle =
\delta (\mathbf{P}'-\mathbf{P}) \langle \mathbf{p}' \Vert T(z) 
\Vert \mathbf{p} \rangle ,
\label{d.3}
\eeq
where the reduced matrix elements $ \langle \mathbf{p}' \Vert V \Vert
\mathbf{p} \rangle$ and $\langle \mathbf{p}' \Vert T(z) \Vert
\mathbf{p} \rangle$ are matrices in spin-isospin space.

Any operator $V$ on the two-nucleon Hilbert space can be expanded
as a linear combination of the spin-basis operators, $
\Sigma_{\mu \nu}:= \sigma_{1\mu} \otimes \sigma_{2 \nu}$,
where $\sigma_{i\mu} = (I,\pmb{\sigma})$, 
\beq
V = \sum V^{\mu\nu} \Sigma_{\mu \nu}
\eeq
and
\beq
V^{\mu\nu} = \frac{1}{4} Tr(\Sigma_{\mu \nu}  V).
\eeq
The operator $\Sigma_{\mu \nu}$ has sixteen components, however the
number of independent operator types that can appear in a transition
operator or potential is constrained by rotational invariance,
time-reversal invariance, space-reflection symmetry, and spin-exchange
symmetry.  Spin-exchange symmetry means that the potential commutes with the
square of the total spin, which is a symmetry of nucleon-nucleon 
interactions.  These symmetries can be utilized to reduce number of coupled 
Lippmann-Schwinger equations. 

The three Pauli spin matrices in $\Sigma_{\mu \nu}$ for each particle
can be replaced by the rotationally invariant operators,
$\mathbf{\hat{{V}}_i}\cdot \pmb{\sigma}_j$, for any independent set of
vector operators,
$\{\mathbf{\hat{{V}}_1},\mathbf{\hat{{V}}_2},\mathbf{\hat{{V}}_3} \}$.
The traces of the product of these operators with the potential have
the same time-reversal, space-reflection, and spin-exchange symmetry
as the operators.  If the spin operators are chosen to be orthonormal
with respect to the trace norm, the only operators that contribute to
the potential expansion are the operators that satisfy all these
symmetries, except time-reversal invariance, because it is possible to
have a scalar coefficient function that changes sign on time reversal.
Consideration of the symmetries of a complete set of 
spin operators implies that the most general interaction consistent
with these symmetries can be represented by an expansion in terms of
five operators on shell and six off shell.  When the spin operators
are not orthonormal with respect to the trace, a matrix inversion is
needed to compute the expansion of the potential.  This inverse
matrix may not have any simple symmetry with respect to time
reversal.  The same will be true for the resulting expansion
coefficients.  Nevertheless, the number of operators needed is
unchanged and the symmetries of the full potential are preserved.

In this paper we consider the following
independent momentum operators used by Wolfenstein~\cite{wolfenstein}
\begin{eqnarray}
\hat{\mathbf{K}} &:=& {(\mathbf{p}'-\mathbf{p}) \over \vert 
\mathbf{p}'-\mathbf{p}\vert} \cr
\label{d.4}
\hat{\mathbf{Q}}& :=& {(\mathbf{p}'+\mathbf{p}) \over \vert 
\mathbf{p}'+\mathbf{p}\vert} \cr
\label{d.5}
\hat{\mathbf{N}} &:=& {(\mathbf{p}' \times\mathbf{p}) \over \vert 
\mathbf{p}'\times \mathbf{p}\vert}.
\label{d.6}
\end{eqnarray}
These vector operators can be classified by their transformation
properties with respect to space reflection and time reversal.

A useful independent set of spin operators that are sufficient to 
expand the scattering amplitude matrix $M$ was given by
Wolfenstein.  
In terms of the vectors of Eq.~(\ref{d.6}) the Wolfenstein operators are 
\beq
\Big\{ I, 
(\bsigma_1 \cdot \hat{\mathbf{Q}}) \otimes 
(\bsigma_2 \cdot \hat{\mathbf{Q}}) ,
(\bsigma_1 \otimes  I  + 
I \otimes \bsigma_2)\cdot \hat{\mathbf{N}},
(\bsigma_1 \cdot \hat{\mathbf{N}}) \otimes 
(\bsigma_2 \cdot \hat{\mathbf{N}}) ,
(\bsigma_1 \cdot \hat{\mathbf{K}}) \otimes 
(\bsigma_2 \cdot \hat{\mathbf{K}}) \Big\} .
\label{d.7}
\eeq
The coefficients of the expansion of the scattering amplitude matrix
in Wolfenstein spin-operators are the so-called Wolfenstein parameters,
$a,c,m,g,h$.  Knowledge of these parameters as a function of the
initial and final momenta contains all of the information in the
scattering amplitude matrix. The scattering-amplitude matrix 
is related to the on-shell transition matrix element by 
\beq
M (\mathbf{p}', \mathbf{p}) = 
- 4 \pi^2 \mu \langle \mathbf{p}' \Vert T(z) \Vert \mathbf{p} \rangle 
\label{d.8}
\eeq
where $\mu$ is the reduced mass of the two-nucleon system 
and the Wolfenstein parameterization of $M (\mathbf{p}', \mathbf{p}) $ is 
\begin{eqnarray}
M&=& a I + c (\bsigma_1 \otimes  I  + I \otimes \bsigma_2)\cdot \hat{\mathbf{N}} + 
m (\bsigma_1 \cdot \hat{\mathbf{N}}) \otimes (\bsigma_2 \cdot \hat{\mathbf{N}}) + \cr
& & (g+h)(\bsigma_1 \cdot \hat{\mathbf{Q}}) \otimes 
(\bsigma_2 \cdot \hat{\mathbf{Q}}) +
(g-h)(\bsigma_1 \cdot \hat{\mathbf{K}}) \otimes 
(\bsigma_2 \cdot \hat{\mathbf{K}}) .
\label{d.9}
\end{eqnarray}
The operators of Eq.~(\ref{d.9}) are invariant with respect to rotations,
space reflection, spin exchange, and time reversal.

The additional operator,
\beq
(\bsigma_1 \cdot \hat{\mathbf{K}}) \otimes 
(\bsigma_2 \cdot \hat{\mathbf{Q}}) +
(\sigma_1 \cdot \hat{\mathbf{Q}}) \otimes 
(\bsigma_2 \cdot \hat{\mathbf{K}}), 
\label{d.10}
\eeq
is rotationally invariant, space reflection invariant and spin
exchange invariant, but changes sign under time reversal.  If this operator is
paired with a coefficient function that also changes sign under time
reversal then the product is invariant with respect to time reversal.
A time-odd coefficient function must be odd in $\vert \mathbf{p}' \vert^2 - \vert
\mathbf{p} \vert^2 $, and thus vanish on shell.  This means that while this
operator cannot appear in the scattering amplitude matrix, $M$, it
can appear in the interaction or half-shell transition matrix and it
may also appear in the off-shell unitarity constraint on the
transition operator.

The six independent operators of Eqs.~(\ref{d.7}) and (\ref{d.10}) 
are sufficient to expand any
potential that is invariant with respect to
rotations, space reflection, time reversal, and spin exchange.

In order to solve the Lippmann-Schwinger equation the interaction is
expanded in terms of these six independent spin operators.  In our
applications the operator $(\bsigma_1 \cdot \hat{\mathbf{N}}) \otimes (\bsigma_2
\cdot \hat{\mathbf{N}})$ is replaced by the operator $(\bsigma_1 \cdot \bsigma_2)
$.  This replacement is done for numerical reasons, but it is also useful
because this operator appears in the spin-spin part of the interaction
as well as in the tensor force.
 
The transition matrix elements can be expanded using the same six
operators.  This results in a set of six coupled equations for the
half-shell or off-shell transition matrix elements.  Since the
scattering amplitude matrix is related to the on-shell transition
operator by a multiplicative scalar, it is useful to put the integral
equation in a form where the on-shell points are included in the
linear equation.  To avoid ambiguities, the on-shell amplitude is
expanded using 5 independent spin components.  This will be discussed
in section IV.

The expansion of both the interactions and transition matrix elements is  
constructed by taking traces with the six independent spin operators
\begin{eqnarray}
W^1& :=& {\bf I} \cr 
W^2 &:= & \bsigma_1 \cdot \bsigma_2 \cr
W^3 &:=&
(\bsigma_1 \cdot \hat{\mathbf{K}}) \otimes (\bsigma_2 \cdot \hat{\mathbf{K}}) \cr
W^4& :=& (\bsigma_1 \cdot \hat{\mathbf{Q}}) \otimes 
(\bsigma_2 \cdot \hat{\mathbf{Q}}) \cr
W^5&:=& i (\bsigma_1 \otimes  {\bf I}  + 
{\bf I} \otimes \bsigma_2)\cdot \hat{\mathbf{N}} \cr
W^6&:=& (\bsigma_1 \cdot \hat{\mathbf{K}}) \otimes 
(\bsigma_2 \cdot \hat{\mathbf{Q}}) +
(\bsigma_1 \cdot \hat{\mathbf{Q}}) \otimes 
(\bsigma_2 \cdot \hat{\mathbf{K}}) . 
\label{d.16}
\end{eqnarray}
It follows that 
\beq
V = \sum_{i=1}^6 U_i W^i ,
\label{d.17}
\eeq
where
\beq
U_i := (A^{-1})^{ij} \mathbf{Tr}(W^j V)
\qquad
A^{ij} := \mathbf{Tr}(W^i W^j). 
\label{d.18}
\eeq
Both $T(z)$ and $V$ are matrices in the spin and isospin degrees
of freedom and the $W^j$ and $A^{ij}$ depend on $\mathbf{p}$ and
$\mathbf{p}'$.  By taking traces, the Lippmann-Schwinger
equation becomes a set of six coupled integral equations in 
two variables.

As discussed before, six linearly independent operators are required
to expand any NN potential. 
This expansion can include the product of a time-odd, space reflection
invariant spin-momentum operator with a space reflection invariant,
time-odd coefficient function resulting in a term that conserves the
symmetry properties of the NN-potential.  Note that a space-reflection
odd operator cannot appear in this expansion because it would require
a pseudo-scalar coefficient which cannot be constructed from two
vectors~\cite{saravanan-thesis}.  

The spin-spin operator, $W^2$, and the Wolfenstein operator 
$(\bsigma_1\cdot{\bf N}) (\bsigma_2\cdot{\bf N})$ are independent and 
have all of symmetries of the potential.  They are related by
\begin{eqnarray}
\lefteqn{\bsigma_1 \cdot\bsigma_2=} \cr
& & R\left(\frac{R^{2}}{{( K^2 Q^2-R^2)}^{2}}-\frac{ K^{2} Q^{2}}{{( K^{2}
Q^{2}-R^{2})}^{2}}\right) \;
\left( (\bsigma_1 \cdot {\bf K}) \; (\bsigma_2 \cdot {\bf Q} )  + 
(\bsigma_1 \cdot {\bf Q})  (\bsigma_2 \cdot {\bf K}) \right) \cr
&+& \left(\frac{ Q^{2} R^{2}}{{( K^{2} Q^{2}-R^{2})}^{2}}-\frac{ K^{2} Q^{4}}{{( K^{2}
Q^{2}-R^{2})}^{2}}\right) \; \left(\bsigma_1 \cdot {\bf K}\right) 
\left(\bsigma_2\cdot{\bf K}\right) +  
\frac{ (\bsigma_1\cdot{\bf N}) (\bsigma_2\cdot{\bf N})}{N^{2}} \cr
&+&  \left(\frac{ K^{2} R^{2}}{{( K^{2} Q^{2}-R^{2})}^{2}}-\frac{ K^{4} Q^{2}}{{( K^{2}
Q^{2}-R^{2})}^{2}}\right) \left(\bsigma_1\cdot{\bf Q}\right) 
\left(\bsigma_2\cdot{\bf Q}\right),
\label{id4}
\end{eqnarray}
which involves $W^3$, $W^4$ as well as the time-odd operator, $W^6$.  
In this expression $R={\bf K}\cdot{\bf Q}=\vert\mathbf{p}'\vert^{2}-\vert \mathbf{p}\vert^{2}$ is the time-odd 
coefficient of the time-odd operator $W^6$, which vanishes on-shell.

It follows that the basis for the off-shell potential, $W^1-W^6$ could
be replaced by an equivalent basis where $W^6$ is replaced by the
Wolfenstein operator $(\bsigma_1\cdot{\bf N}) (\bsigma_2\cdot{\bf
N})$.  While the second basis has all of the symmetries of the 
potential,  the basis that includes the time-odd operator $W^6$
is preferable for numerical reasons as will be elaborated on later.

\section{Lippmann-Schwinger Equations}

The reduced transition matrix elements can be expressed 
as
\beq
t(p',p,x; z) := 
\langle \mathbf{p}' \Vert T(z) \Vert \mathbf{p} \rangle  
 \qquad  z := \frac{p_0^2}{2 \mu} + i 0^+ ,
\label{e.2}
\eeq
where the variables are defined below and    
$t(p',p ,x; z)$ is a spin-isospin valued function.
Here $p_0$ represents the on-shell momentum. 
The Lippmann-Schwinger equation for $t(p',p,x; z)$ is
the two-variable integral equation
\begin{eqnarray}
\lefteqn{ t(p',p, x'; z) =} \cr 
& &  v(p',p, x') + 
\int_{-1}^1 dx'' \int_0^\infty  dp'' 
p^{\prime\prime 2}   u(p',x', p'',x'' )G_0 (p'',z) 
t(p'',p,x''; z) \; ,
\label{e.3}
\end{eqnarray}
where the coordinate system is chosen so that the initial momentum is in
the $3$-direction and the scattering plane is the $1$-$3$ plane. 
The integration over the azimuthal angle only affects variables in the potential and can be 
 carried out independently. 
The momenta and quantities in Eq.~(\ref{e.3}) are parameterized in terms 
of the variables $p, p', x' , p'', x''$ defined by
\begin{eqnarray}
\mathbf{p} & :=&  (0,0,p)=\mathbf{p}_0 \cr
\label{e.4}
\mathbf{p}'&:=& \left(p' \sqrt{1-x^{\prime 2}},0,p'x'\right) \cr
\label{e.5}
\mathbf{p}''&:= & \left(p'' \sqrt{1-x^{\prime \prime 2}}\cos (\phi''),
p'' \sqrt{1-x^{\prime \prime 2}}\sin (\phi''),
p'' x''\right)  \cr
\label{e.6}
\hat{\mathbf{p}}' \cdot \hat{\mathbf{p}} & :=&  x' \cr
\label{e.7}
\hat{\mathbf{p}}'' \cdot \hat{\mathbf{p}} &:=& x'' \cr
\label{e.8}
\hat{\mathbf{p}}' \cdot \hat{\mathbf{p}}''& :=&  y =
x' x'' + \sqrt{(1-x^{\prime 2})(1-x^{\prime \prime 2})}\cos (\phi'') \cr
\label{e.9} 
v (p',p,x') &:= &V(\mathbf{p}',\mathbf{p})   \cr
\label{e.10}
v (p',p'',y)  &:= & V(\mathbf{p}',\mathbf{p}'')   \cr
\label{e.10a}
 u(p',x',p'',x'',\phi'') &:=&  v (p',p'',x' x'' +
\sqrt{(1-x^{\prime 2})(1-x^{\prime \prime 2})}\cos (\phi''))  \cr
\label{e.11}
G_0 (p'',z) &=& {2 \mu \over p_0^2 - p^{\prime\prime^2}+i0^+}   \cr
u(p',x',p'',x'')&:=& \int_0^{2 \pi} d\phi'' u(p',x',p'',x'',\phi'').
\label{e.12}
\end{eqnarray}
In the following we omit the explicit limits of the integrals.
One can see that Eq.~(\ref{e.3}) is an integral equation in two variables, the
integration over the azimuthal angle $\phi''$ can be carried out independently
as given in Eq.~(\ref{e.12}). 

In order to reduce Eq.~(\ref{e.3}) to an algebraic equation the 
singular integral is first treated by a subtraction
\begin{eqnarray}
\lefteqn{t(p',p,x';z) = v(p',p,x') +} \cr
&\int& dx''  dp'' {2 \mu  \over p_0^2 - p^{\prime \prime 2} } 
\left[ u(p',x',p'',x'') p^{\prime\prime 2} t(p'',p,x'';z)-
u(p',x',p_0,x'') p_0^{2} t(p_0,p,x'';z)\right] \cr
& & - {i 2 \mu \pi p_0 } \int 
dx'' u(p',x',p_0,x'') \; t(p_0,p,x'';z) \; ,
\label{e.13}
\end{eqnarray}
where the unknown $t(p_0,p,x'';z)$ satisfies
\begin{eqnarray}
\lefteqn{t(p_0,p,x';z) = v(p_0,p,x') + } \cr
&\int& dx''  dp'' {2 \mu  \over p_0^2 - p^{\prime \prime 2} } 
\big[u(p_0,x',p'',x'') p^{\prime\prime 2} t(p'',p,x'';z)-
u(p_0,x',p_0,x'') p_0^{2} t(p_0,p,x'';z)\big] \cr
& & - {i 2 \mu  p_0 \pi } \int 
dx'' u(p_0,x',p_0,x'') \; t(p_0,p,x'',z) . 
\label{e.14}
\end{eqnarray}
These equations define a non-singular set of coupled integral
equations.  The integrals are approximated by sums over quadrature
points and weights.  The quadrature points are chosen to not include
the point $p_0$, which is treated separately.  The resulting linear
system is given by
\begin{eqnarray}
\lefteqn{t(p_i',p,x_l';z) = v(p_i',p,x_l') +} \cr
& &\sum_{kj} dx_k''  dp_j''  {2 \mu  \over p^2_0 - p_j^{\prime \prime 2} } 
\left[ u(p_i',x_l',p_j'',x_k'') p_j^{\prime\prime 2} 
t(p_j'',p,x_k'';z)-
u(p_i',x_l',p_0,x_k'') p_0^{2} t(p_0,p,x_k'';z)\right]  \cr
& & - {i 2 \pi \mu  p_0 } \sum_k 
dx_k'' u(p',x',p_0,x_k'') t(p_0,p,x_k'';z) 
\label{e.15}
\end{eqnarray}
and
\begin{eqnarray}
\lefteqn{t(p_0,p,x_l';z) = v(p_0,p,x_l') + } \cr
& &\sum_{kj} dx_k''  dp_j''  {2 \mu  \over p_0^2 - p_j^{\prime \prime 2} } 
\left[u(p_0,x_l',p_j'',x_k'') p_j^{\prime\prime 2} t(p_j'',p,x_k'';z)-
u(p_0,x_l',p_0,x_k'',) p_0^{2} t(p_0,p,x_k'';z)\right] \cr
& &- {i 2 \pi \mu  p_0  } \sum_k 
dx''_k u(p_0,x_l',p_0,x_k'') t(p_0,p,x_k'';z). 
\label{e.16}
\end{eqnarray}
This linear system gives approximate solutions for $t(p_i',p,x'_l;z)$
at the quadrature points and at the point where $p_i'=p_0$.

Given the approximate solutions at the quadrature points, the results for any
values of $p',p,x'$ can be computed by inserting the solutions at the
quadrature points back in the integral equation \cite{sloan},
\begin{eqnarray}
\lefteqn{t(p',p,x';z) = v(p',p,x') + } \cr
& &\sum_{jk} dx_k''  dp_j'' {2 \mu  \over p_0^2 - p_j^{\prime \prime 2} } 
\left[ u(p',x',p_j'',x_k'') p_j^{\prime\prime 2} t(p_j'',p,x_k'';z)-
u(p',x',p_0,x_k'' ) p_0^{2} t(p_0,p,x_k'';z) \right] \cr
& &- {i 2 \pi \mu p_0 } \sum_k 
dx_k''u(p',x',p_0,x_k'') t(p_0,p,x_k'';z) .
\label{e.17}
\end{eqnarray}
These equations give the off-shell transition matrix elements;
half-shell matrix elements are obtained by setting $p^2=p_0^2$, while
on-shell matrix elements have $p^2= p^{\prime 2}=p_0^2$.

When calculating NN observables, it is sufficient to solve for the half-shell transition matrix.
This solution can be used as
input \cite{Keister:2005eq,Lin:2008sy}, using the first resolvent
equation \cite{simon1}, to find the off-shell transition matrix.  For
this reason we only discuss the solution of the half-shell equation,
$p^2=p_0^2$.

For realistic models the interactions and transition operators in the
above equations are matrices in the spin and isospin degrees of
freedom.  To treat this we expand $t(p_j',p,x_k';z)$ as a linear
combination of known spin operators and unknown coefficients
functions.  A similar expansion is used to represent the interaction
except the coefficient functions are known for the interactions.  The
resulting equations are for the unknown coefficient functions.
The explicit form of the algebraic equations for the transition matrix elements
are given in Appendix~\ref{appendixa}.

\section{Calculation of the Wolfenstein Amplitudes}

\subsection{Formal Considerations}

As noted in Ref.~\cite{Golak:2010wz}, one of the major challenges in
the calculation of NN observables is the calculation of the on-shell
amplitudes.  In order to avoid taking on-shell limits of off-shell
amplitudes expressed as linear combinations of six independent
spin-operators, we directly calculate the on-shell transition
amplitudes by expanding them using the five linearly independent
Wolfenstein operators given by Eq.~(\ref{d.7}).  The
spin-spin operator in the NN-potential is on-shell linearly dependent on
the Wolfenstein operators via Eq.~(\ref{id4}), since here $R=0$,
\begin{eqnarray}
\bsigma_1\cdot\bsigma_2=\frac{(\bsigma_1 \cdot{\bf K}) \; (\bsigma_2 \cdot{\bf K})}{K^{2}} 
+ \frac{(\bsigma_1\cdot{\bf Q}) \; (\bsigma_2\cdot{\bf Q})}{Q^{2}} 
+\frac{(\bsigma_1 \cdot{\bf N});(\bsigma_2\cdot{\bf N})}{N^{2}} ,
\label{id1}
\end{eqnarray}
where we used the unnormalized vectors, {\bf K}, {\bf Q}, and {\bf N}.
The normalization factors are 
\beq
K^2=2p^2_0 (1-x),  \quad Q^{2}=2p^{2}_0(1+x), \quad N^{2}=p^{4}_0(1-x^{2}),
\eeq
with $x:= \hat{\mathbf{p}} \cdot \hat{\mathbf{p}}'$ and the magnitude
of the on-shell momentum $p_0$.
For numerical computations it is essential to avoid zeros in a denominator.
Thus we re-express 
the operator $(\bsigma_1\cdot{\bf N}) (\bsigma_2\cdot{\bf N})$ as
\begin{eqnarray}
\lefteqn{(\bsigma_1\cdot{\bf N})\; (\bsigma_2\cdot{\bf N}) =} \cr
& & p_0^{4}(1-x^{2}) \; \bsigma_1\cdot\bsigma_2 - \frac{p_{0}^{2}(1-x)}{2}
(\bsigma_1\cdot {{\bf Q}})(\bsigma_2\cdot {{\bf Q}})-\frac{p_{0}^{2}(1+x)}{2}
(\bsigma_1\cdot {{\bf K}})(\bsigma_2\cdot{{\bf K}}),
\label{id3}
\end{eqnarray}
which is well behaved in the limits
$x \rightarrow +/- 1$ and $p_0 \rightarrow 0$. 
This justifies our choice to use Eq.~(\ref{id3}) to replace the 
operator $(\bsigma_1\cdot{\bf N}) (\bsigma_2\cdot{\bf N})$ 
in the Wolfenstein basis with the spin-spin operator 
$W^2= \bsigma_1\cdot\bsigma_2$. 

For the sixth operator we considered two choices: 
One of them
is the time-odd operator $W^6 = ((\bsigma_1\cdot{\bf K}) \;
(\bsigma_2\cdot{\bf Q}) + (\bsigma_1\cdot{\bf Q})(\bsigma_2\cdot{\bf K}))$.
The coefficient function in the NN potential for this operator vanishes
on-shell.

A second choice is the Wolfenstein operator 
$(\bsigma_1\cdot{\bf N})\; (\bsigma_2\cdot{\bf N})$~\cite{Golak:2010wz}.
This operator is related to 
the time-odd operator, $W^6$, by re-expressing 
Eq.~(\ref{id4}) as
\begin{eqnarray}
\lefteqn{(\bsigma_1\cdot{\bf N})(\bsigma_2\cdot{\bf N})=} \cr
& &N^{2}(\bsigma_1\cdot\bsigma_2) + \left(\frac{Q^{2} }{4}\right)
(\bsigma_1\cdot{\bf K})
(\bsigma_2\cdot{\bf K})
 +\frac{({\bf K}\cdot{\bf Q})}{4}
\left ((\bsigma_1\cdot{\bf K})(\bsigma_2\cdot{\bf Q}) + (\bsigma_1\cdot{\bf Q})
(\bsigma_2\cdot{\bf K})\right ) \cr
&+&  \left(\frac{K^{2}}{4}\right) (\bsigma_1\cdot{\bf Q})
(\bsigma_2\cdot{\bf Q}). 
\end{eqnarray}
If we choose the sixth operator to be 
$(\bsigma_1\cdot{\bf N})\;(\bsigma_2\cdot{\bf N})$,
 the on- and off-shell potentials coefficient 
functions are related by
\begin{eqnarray}
V_{1}^{\rm{onshell}}&=&V_{1}^{\rm{offshell}} \cr
V_{2}^{\rm{onshell}}&=&V_{2}^{\rm{offshell}} -  p_0^{4}(1-x^{2})V_{6}^{\rm{offshell}} \cr 
V_{3}^{\rm{onshell}}&=&V_{3}^{\rm{offshell}}-\frac{p_{0}^{2}(1+x)}{2}V_{6}^{\rm{offshell}} \cr
V_{4}^{\rm{onshell}}&=&V_{4}^{\rm{offshell}}-\frac{p_{0}^{2}(1-x)}{2}V_{6}^{\rm{offshell}} \cr
V_{5}^{\rm{onshell}}&=&V_5^{\rm{offshell}} \;,
\label{choice1}
\end{eqnarray}
where $V_{i}$ are the expansion functions for the operators $i={1,2,3,4,5}$. 
These expressions show a rapid variation of these coefficients near the 
on-shell point, making the direct computation of the on-shell values in Eq.~(\ref{e.17}) essentially impossible.

However, when the time-odd operator, $W^6$, is chosen as 6th operator,
the relationship between the on-shell and off-shell coefficient functions is 
\begin{eqnarray}
V_{1}^{\rm{onshell}}&=&V_{1}^{\rm{offshell}} \cr
V_{2}^{\rm{onshell}}&=&V_{2}^{\rm{offshell}} \cr
V_{3}^{\rm{onshell}}&=&V_{3}^{\rm{offshell}} \cr
V_{4}^{\rm{onshell}}&=&V_{4}^{\rm{offshell}} \cr
V_{5}^{\rm{onshell}}&=&V_{5}^{\rm{offshell}} ,
\label{choice2} 
\end{eqnarray}
which does not exhibit rapid variations near
the on-shell point. Thus, calculations with the time-odd operator as 
part of the basis have the advantage that the coefficient functions are on- and off-shell the same. 
Test calculations using
$(\bsigma_1\cdot{\bf N})\; (\bsigma_2\cdot{\bf N})$ as 6th
operator led to numerical instabilities in  solving the
Lippmann-Schwinger equation, as already pointed out in Ref.~\cite{Golak:2010wz}.
Thus, our calculations use the 
time-odd operator, $W^6$,  as the 6th operator in the basis.

The scattering amplitude matrix can be expressed in terms of the 
on-shell transition matrix elements \cite{glockle:1983} as 
\beq
M (\mathbf{p}', \mathbf{p}) = 
- 4 \pi^2 \mu t(p_0,p_0,x,z)
=- 4 \pi^2 \mu \sum_{i=1}^{5}t_{i}(p_0,p_0,x,z)W^{i}(p_0,p_0,x),
\label{g.1}
\eeq
where we do not display the spin-isospin parameters.

Accounting for all of the invariances, $M$ can be expressed in the 
basis of operators
of Eq.~(\ref{d.9}) in terms of the Wolfenstein parameters 
$\{a,c,m,g,h\}$.  These complex coefficients encode all of the
information contained in the on-shell transition matrix elements.
By writing the Wolfenstein operators in term of the 
operators of Eq.~(\ref{d.16}), the  matrix $M$
can be expressed in terms of the Wolfenstein parameters as
\beq
M = a W^1 + m W^2 + (g-h-m ) W^3 + (g+h-m) W^4 -i c W^5 .
\label{g.2.1}
\eeq

For identical nucleons this expression has to be anti-symmetric under
the exchange of the nucleons,
including the isospin factors.
This can be achieved by anti-symmetrizing the 
initial two-nucleon state.  In our operator formalism
a particle interchange is represented by  
a spin exchange and the reversal of the initial momentum,
$\mathbf{p} \to - \mathbf{p}$.  The spin-exchange operator is given by
\beq
P_{12} = \frac{1}{2} \left(I + \pmb{\sigma}^1\cdot \pmb{\sigma}^2\right)
= \frac{1}{2} \left(I + W^2\right).
\eeq
If we change the sign of the initial momentum  $\mathbf{p}$, the vectors 
$\mathbf{K}, \mathbf{Q}$, and $\mathbf{N}$ become 
\begin{eqnarray}
\mathbf{K} &\to& \mathbf{Q}  \cr
\mathbf{Q}& \to& \mathbf{K}  \cr
\mathbf{N}& \to& -\mathbf{N} ,
\end{eqnarray}
which implies that the spin-basis elements transform as
\begin{eqnarray}
W^1 & \to  &W^1  \cr
W^2  &\to  &W^2  \cr
W^3  &\to  &W^4  \cr
W^4  &\to  &W^3  \cr
W^5  &\to  &-W^5  \;.
\end{eqnarray}
The coefficients of the exchange matrix elements become 
\begin{eqnarray}
W^1 &\to &  W^1 P_{12} = {1 \over 2} W^1 ( I+ W^2) = {1 \over 2} ( W^1+ W^2) \cr
W^2 &\to & W^2 P_{12} = {1 \over 2} W^2 ( I+ W^2) = {1 \over 2} (3 W^1 -W^2) \cr
W^3 &\to & W^4 P_{12} = {1 \over 2} W^4 ( I+ W^2) =  W^4 + 
{Q^2 \over 2}   (W^1-W^2)  \cr
W^4 &\to & W^3 P_{12} = {1 \over 2} W^3 ( I+ W^2) = W^3 + { K^2 \over 2} 
(W^1-W^2)  \cr
W^5 &\to & -W^5 P_{12} = - {1 \over 2} W^5 ( I+ W^2) =-  W^5 . 
\end{eqnarray}
The expansion coefficients of the $W^i$ become 
\beq
t_{i}(p_0,p_0,x) \to t_{i}(p_0,p_0,-x).
\eeq
Putting all of this together using $t=1$ to denote the symmetric 
iso-triplet subspace,
$t=0$ to denote the antisymmetric iso-singlet subspace, 
$m_i (\mathbf{p}',\mathbf{p}) := - 4 \pi^2  \mu t_{i}(p_0,p_0,x)$
and 
$m_i (\mathbf{p}',-\mathbf{p}) = - 4 \pi^2  \mu t_{i}(p_0,p_0,-x)$
we get the following expansion for scattering amplitude matrix 
including exchange contributions
\begin{eqnarray}
\lefteqn{M=
W^1 \left( m_1 (\mathbf{p}',\mathbf{p}) +
{1 \over 2} (-1)^t m_1 (\mathbf{p}',-\mathbf{p}) +
{3 \over 2} (-1)^t m_2 (\mathbf{p}',-\mathbf{p}) +
\right .}  \cr
 & & +\left .
{1 \over 2} (-1)^t m_3 (\mathbf{p}',-\mathbf{p})+
{1 \over 2} (-1)^t m_4 (\mathbf{p}',-\mathbf{p})
\right)  \cr
 & & + W^{2} \left(  m_2 (\mathbf{p}',\mathbf{p})
+  {1 \over 2} (-1)^t m_1 (\mathbf{p}',-\mathbf{p})
-  {1 \over 2} (-1)^t m_2 (\mathbf{p}',-\mathbf{p}) \right . \cr
& &  \left . - {1 \over 2} (-1)^t  m_3 (\mathbf{p}',-\mathbf{p})
- {1 \over 2} (-1)^t  m_4 (\mathbf{p}',-\mathbf{p})\right)  \cr
 & & + 
W^3 \left( m_3 (\mathbf{p}',\mathbf{p}) +(-)^tm_4 (\mathbf{p}',-\mathbf{p})
\right) \cr
 & &  +
W^4 \left( m_4 (\mathbf{p}',\mathbf{p}) +(-)^tm_3 (\mathbf{p}',-\mathbf{p})
\right)  \cr
& & +
W^5 \left( m_5 (\mathbf{p}',\mathbf{p}) -(-)^tm_5 (\mathbf{p}',-\mathbf{p})
\right).
\end{eqnarray}
Comparing this with Eq.~(\ref{g.2.1}) gives the following expressions for the 
Wolfenstein parameters
\begin{eqnarray}
a & =&  m_1 (\mathbf{p}',\mathbf{p}) +
\frac{1}{2} (-1)^t m_1 (\mathbf{p}',-\mathbf{p}) +
\frac{3}{2} (-1)^t m_2 (\mathbf{p}',-\mathbf{p}) \cr
 & &  + \frac{1}{2} (-1)^t m_3 (\mathbf{p}',-\mathbf{p})+
\frac{1}{2} (-1)^t m_4 (\mathbf{p}',-\mathbf{p})
  \cr
m & =  & m_2 (\mathbf{p}',\mathbf{p})  + 
(-1)^t \frac{1}{2} m_1 (\mathbf{p}',-\mathbf{p})
- \frac{1}{2} (-1)^t m_2 (\mathbf{p}',-\mathbf{p}) \cr
& & - \frac{1}{2} (-1)^t  m_3 (\mathbf{p}',-\mathbf{p})
- \frac{1}{2} (-1)^t  m_4 (\mathbf{p}',-\mathbf{p}) \cr
c& =& i \left ( m_5 (\mathbf{p}',\mathbf{p}) -(-)^tm_5 (\mathbf{p}',-\mathbf{p})
\right ) \cr
g+h & = &  m_2 (\mathbf{p}',\mathbf{p})+   m_4 (\mathbf{p}',\mathbf{p}) +
- \frac{1}{2} (-1)^t m_2 (\mathbf{p}',-\mathbf{p})  \cr
 & & + \frac{1}{2}(-)^t m_3 (\mathbf{p}',-\mathbf{p})
- \frac{1}{2} (-1)^t  m_4 (\mathbf{p}',-\mathbf{p})
+ \frac{1}{2} (-1)^t m_1 (\mathbf{p}',-\mathbf{p}) \cr
g-h &=& m_2 (\mathbf{p}',\mathbf{p}) +  m_3 (\mathbf{p}',\mathbf{p}) -
\frac{1}{2} (-1)^t m_2 (\mathbf{p}',-\mathbf{p}) \cr
 & & +
\frac{1}{2} (-)^t m_4 (\mathbf{p}',-\mathbf{p}) -
\frac{1}{2} (-1)^t  m_3 (\mathbf{p}',-\mathbf{p})
+ \frac{1}{2} (-1)^t m_1 (\mathbf{p}',-\mathbf{p}).
\label{acmgh}
\end{eqnarray}

\subsection{Wolfenstein Amplitudes for the AV18 Potential}
\begin{figure}
\begin{center}
\includegraphics[scale=.65]{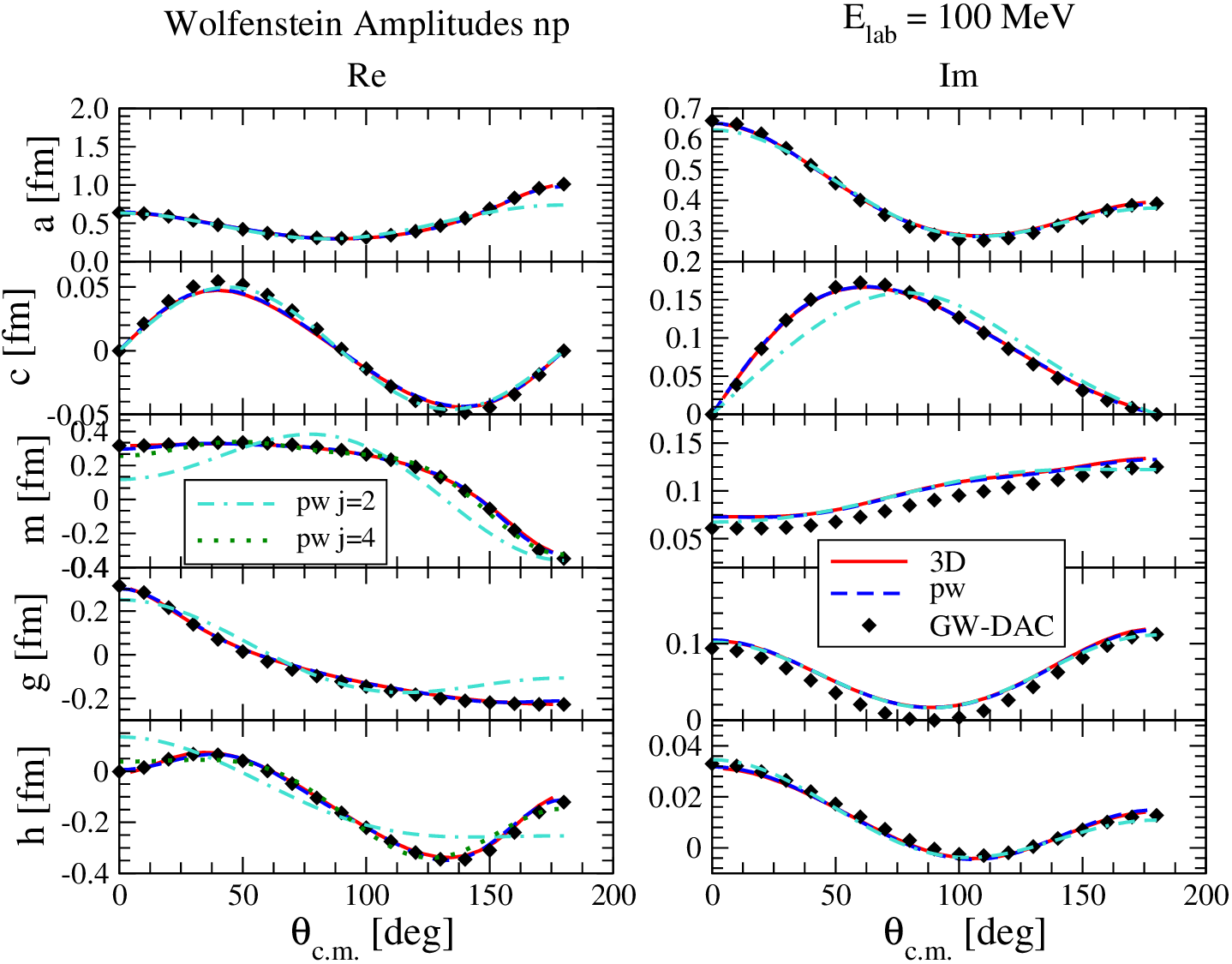}
\caption{The Wolfenstein amplitudes for neutron-proton scattering at 100~MeV
laboratory kinetic energy based on the AV18 potential. The solid (red) line
represents the 3D calculation, whereas the dashed (blue) curve is obtained from
a partial-wave calculation summing partial waves up to $j=6$. The partial-wave
sums up to $j=2$ and $j=4$ are shown as double-dash-dotted (turquoise) and
dotted (green) lines. The data points are the amplitudes extracted from the GW-DAC 
current analysis~\cite{GW-DAC,Arndt:2007qn}.
}
\label{fig1}
\end{center}
\end{figure}

In this subsection we calculate the Wolfenstein amplitudes 
for neutron-proton ($np$) and proton-proton ($pp$) scattering calculated
with our 3D formulation  at 100~MeV and at 350~MeV laboratory energy and 
compare them with calculations~\cite{Weppner:1997wx} 
based on summing partial wave amplitudes up 
to a given $j_{max}$. We chose 100~MeV as the low energy, since here 
already the 
sum over a small number of partial waves should suffice to achieve agreement
with the 3D calculation. This is indeed the case as illustrated in 
Fig.~\ref{fig1} for the
$np$ 
Wolfenstein amplitudes. The solid line represents the 3D calculations, 
which perfectly agrees with a partial-wave calculation in which $j_{max}=6$.  
The figure also indicates that at this energy even $j_{max}=4$ already converges 
to the 3D result. We also compare our calculations to an extraction of the 
$np$ Wolfenstein amplitudes from the current solution of the 
GW-DAC Data Analysis Center~\cite{Arndt:2007qn,Arndt:2008uc,GW-DAC}. 

\begin{figure}
\begin{center}
\includegraphics[scale=.65]{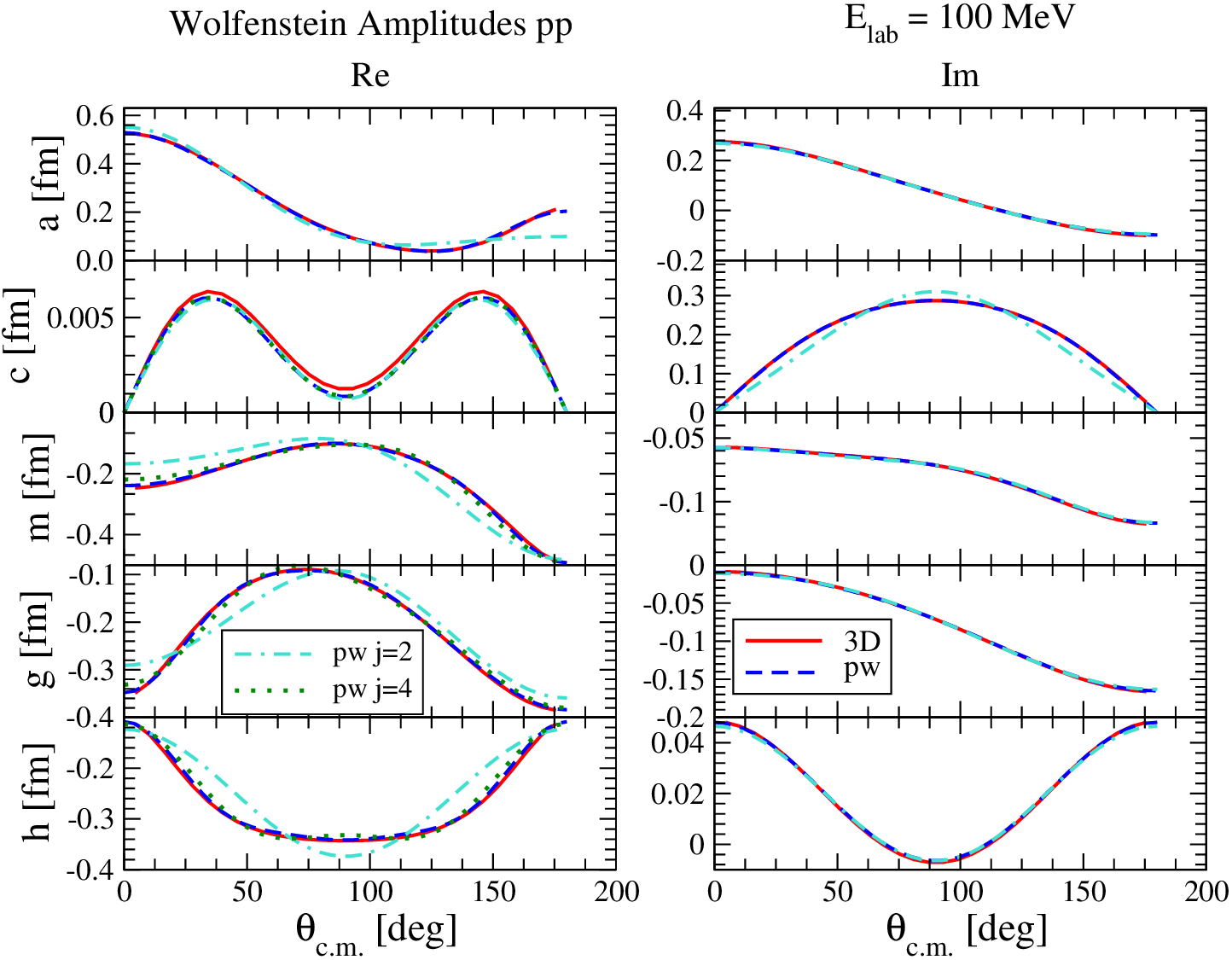}
\caption{The Wolfenstein amplitudes for proton-proton scattering a 100~MeV laboratory kinetic energy based on the AV18 potential. The meaning of the curves is the same as in Fig.~\ref{fig1}.
}
\label{fig2}
\end{center}
\end{figure}

In Fig.~\ref{fig2} the Wolfenstein amplitudes for $pp$ scattering are shown for
100~MeV laboratory scattering energy, and again the 3D calculation is compared
to a partial wave sum up to $j_{max}=6$. Here $pp$ means that we use the strong
$pp$ interaction as given by the AV18 interaction, however do not treat the long-range Coulomb interaction.
The 3D and partial-wave summed calculation agree rather well with each other.  Only the Real $c$ amplitude displays a small deviation. 
However, when considering the scale of the y-axis, this deviation should not be considered 
significant.
In contrast to the $np$ case,  the real part of the amplitudes $g$ and $h$ indicate that for 
very small angles a $j_{max}=4$ calculation is not yet fully converged.  

\begin{figure}
\begin{center}
\includegraphics[scale=.65]{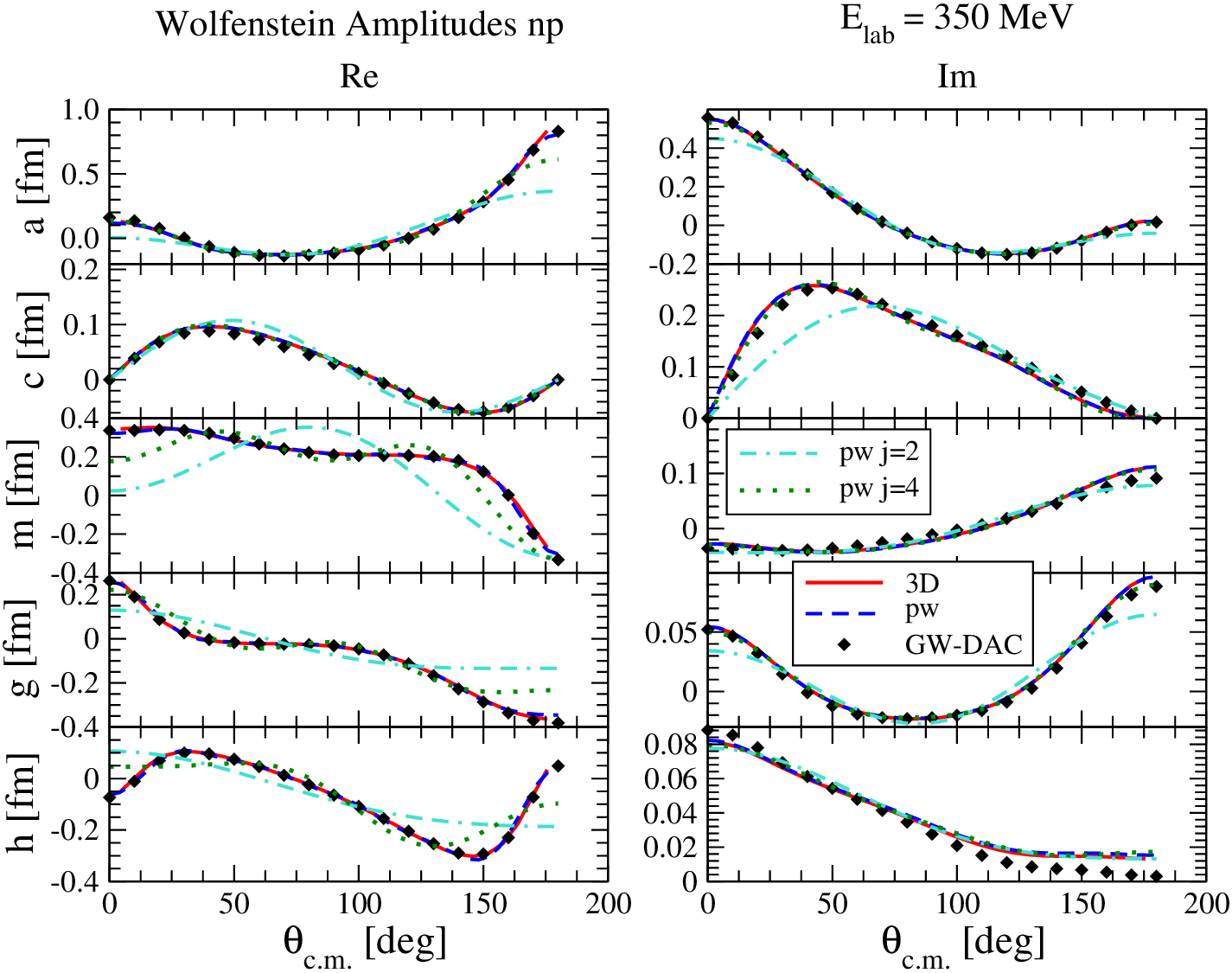}
\caption{The Wolfenstein amplitudes for neutron-proton scattering at 350~MeV
laboratory kinetic energy based on the AV18 potential.The solid (red) line
represents the 3D calculation, whereas the dashed (blue) curve is obtained from
a partial-wave calculation summing partial waves up to $j=9$. The partial-wave
sums up to $j=2$ and $j=4$ are shown as double-dash-dotted (turquoise) and
dotted (green) lines. The data points are the amplitudes extracted from the GW-DAC 
current analysis~\cite{GW-DAC,Arndt:2007qn}. 
}
\label{fig3}
\end{center}
\end{figure}

\begin{figure}
\begin{center}
\includegraphics[scale=.65]{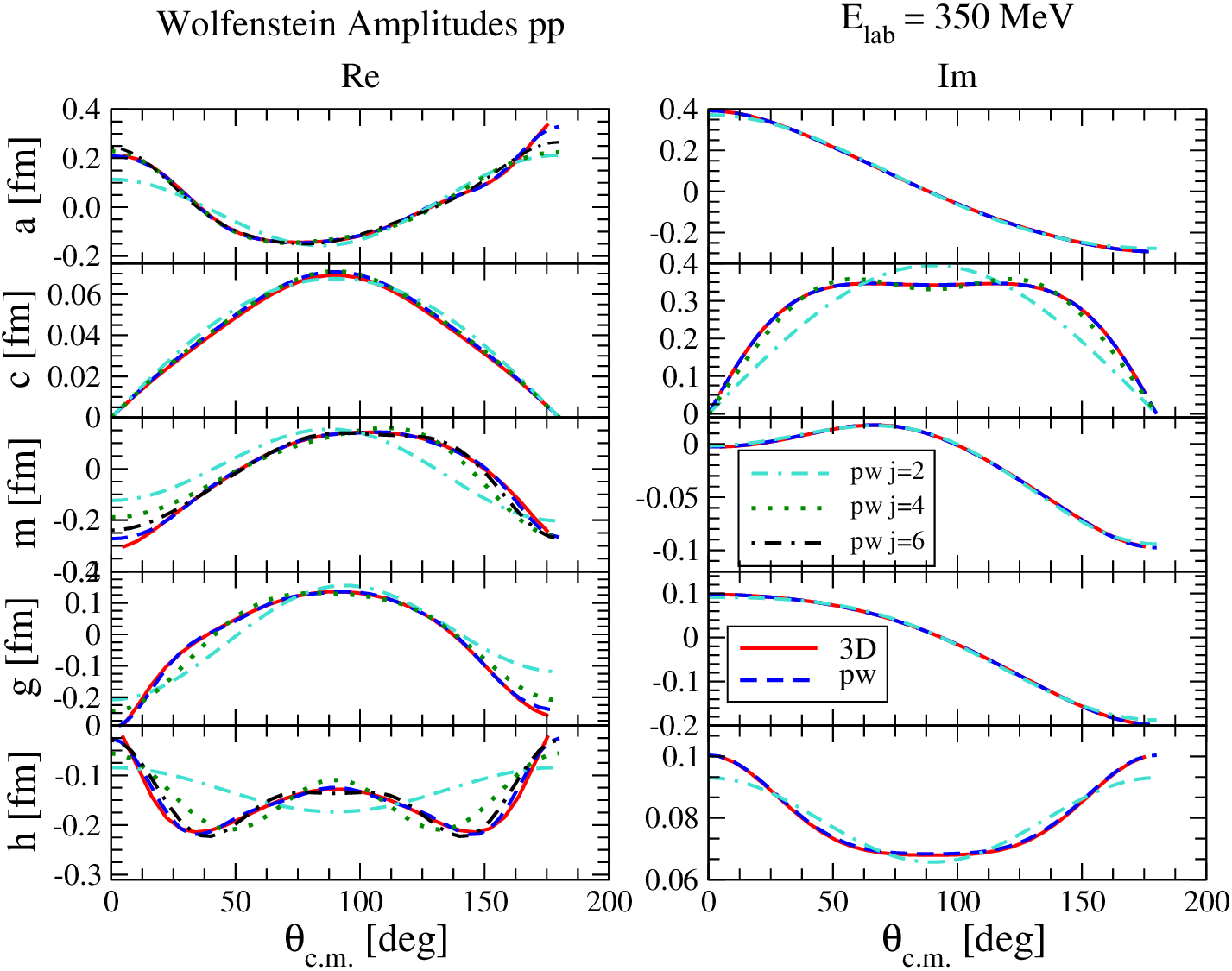}
\caption{The Wolfenstein amplitudes for proton-proton scattering at 350~MeV laboratory kinetic energy based on the AV18 potential. The meaning of the curves is the same as in Fig.~\ref{fig3}.
}
\label{fig4}
\end{center}
\end{figure}

Figs.~\ref{fig3} and \ref{fig4} show the Wolfenstein amplitudes for $np$
and $pp$ scattering at 350~MeV laboratory kinetic energy, the highest
energy for which the AV18 interaction is fitted to the NN data
base. In both figures the solid line represents the 3D calculation,
whereas the dashed line represents a partial-wave calculation summed
up to $j_{max}=9$.  In Fig.~\ref{fig3} the partial-wave sums up to
$j_{max}=2$ and $j_{max}=4$ are also shown. In contrast to the lower
energy of 100~MeV in Fig.~\ref{fig1}, at this energy the partial-wave
sum up to $j_{max}=4$ is clearly not converged. In fact, the backward
angles of the real parts of $a$ and $h$ show a particularly bad
convergence.  In addition, we indicate the extraction of the
Wolfenstein amplitudes from the GW-DAC data
analysis~\cite{Arndt:2007qn,GW-DAC} by the filled diamonds.  For the
$pp$ Wolfenstein amplitudes in Fig.~\ref{fig4} we also display a
partial wave calculation summed up to $j_{max}=6$ to indicate that
even this partial wave sum is not yet converged to the full 3D
calculation. A careful inspection of the real parts of the forward
and backward angles of $a$, $g$, and $h$ reveals that even the partial wave
sum of $j_{max}=9$ is not quite converged yet. In contrast, the
imaginary parts of all amplitudes show already convergence at roughly
$j_{max}=4$.

\subsection{Numerical Details}
\label{erroranalysis}

The LS integral equation that needs to be solved when performing a
calculation without employing a partial-wave decomposition is, in the
form of Eqs.~(\ref{e.13}) and (\ref{e.14}), a two-dimensional integral
equation in two variables (the magnitude of a momentum p' and an angle
x'). Due to the structure of the equation, an angle integration over
the azimuthal angle $\phi$ can be carried out separately as given in
Eq.~(\ref{e.12}), and is thus not part of the integral equation. For
the integration we use standard Gauss-Legendre integration with a
tangent map of the points of the momentum integration. Discretized the set
of coupled equations, Eqs.~(\ref{e.13}) and (\ref{e.14}), turn into a
finite set of linear algebraic equations, which is solved by standard 
methods.

\begin{figure}
\begin{center}
\includegraphics[scale=.65]{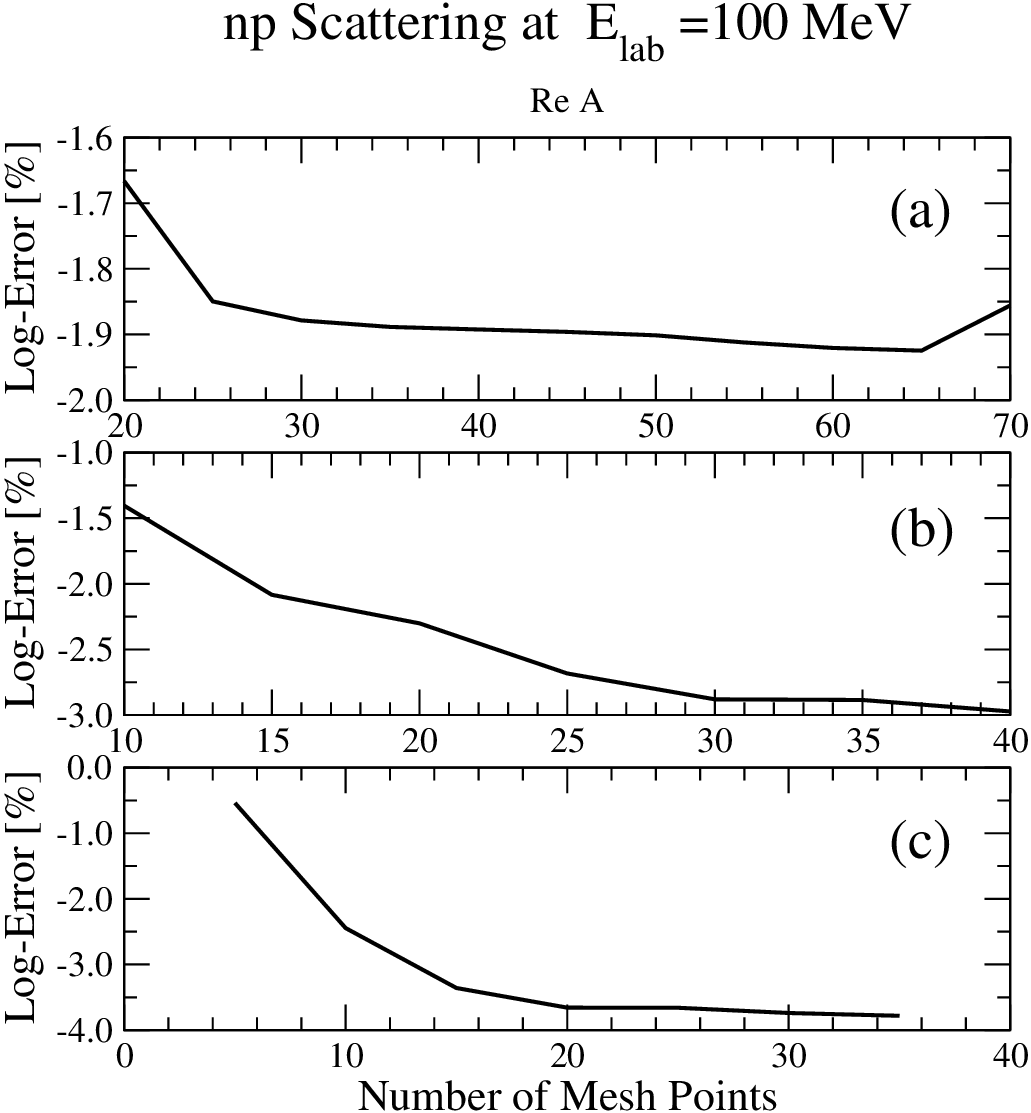}
\caption{The logarithmic error for the momentum grid points (a), the polar angle ($\cos \theta = x$) points (b) and the azimuthal angle ($\phi$) grid points (c) as function of number of points. Precise definitions and explanations are given in Sec.~\ref{erroranalysis}.
}  
\label{fig5}
\end{center}
\end{figure}

However, we need to test the numerical convergence for each
integration. In the following we demonstrate the numerical convergence
by concentrating on the real part of the Wolfenstein $a$ amplitude and
consider its values for different sets of grid points of $p$, $x\equiv
\cos\theta$, and $\phi$.  For the choice of $n$ $p$-points, $m$
$x$-points and $l$ $\phi$-points let us define the value of the real
part of the Wolfenstein amplitude A as $a(n,m,l)$.  Then we define as
logarithmic error in $\phi$,

\beq
{\rm Log-Error}_{\phi} (l)=
\log\left(\left| \frac{(a(60,30,40)-a(60,30,l))*100}{a(60,30,40)}\right|\right) 
\label{eq:errphi}
\eeq 
This Log-Error is plotted in Fig.~\ref{fig5}(c) 
for values $l \in (5,35)$, and we see that the $\phi$ integration converges
very quickly. Already 20 $\phi$ points are sufficient to achieve an error in
the order of $10^{-4}$\%. For most of our calculations we choose 30 $\phi$ points.
For the $x$-points we define
\beq
{\rm Log-Error}_{x}(m)=
\log\left( \left|\frac{(a(60,45,20)-a(60,m,20))*100}{a(60,45,20)}\right|\right)
\label{eq:errx} 
\eeq 
This Log-Error is plotted in Fig.~\ref{fig5}(b) for values $m \in
(10,40)$.  The $x$-integration converges much slower with respect to
the number of mesh-points, and for having an error in the order of
$10^{-3}$to $10^{-4}$\% we need at least 40 integration points
for a converged calculation.

For the momentum-integration points we define
\beq
{\rm Log-Error}_{p}(n)=
\log \left( \left| \frac{(a(75,30,20)-a(n,30,20))*100}{a(75,30,20)}\right|\right)
\label{eq:errp} 
\eeq 
This Log-Error is plotted in Fig.~\ref{fig5}(a) for values $n \in
(20,70)$.  Here we see that the error in the momentum integration
decreases slowest as a function of the momentum-integration points, and
in order to have an error of about $10^{-2}$\% and we choose 60 $p$-points for 
our calculations.  Based on the above considerations, all
of our calculations use $(n,l,m)=(30,40,60)$ mesh points to achieve an
overall numerical errors of 0.01\%, which is well below the accuracy
with which NN observables are given.

\section{Observables}

One of the main purposes of this paper it to test the momentum-space
potential AV18 potential given in~\cite{veerasamy:2100}. There it was
demonstrated that the potential reproduced the deuteron
binding energy and the wave functions. However, these tests were 
only sensitive to the $np$ part of the potential.
Now we have the opportunity to test the $np$ as well as the $pp$ pieces
of the AV18 potential given in Ref.~\cite{veerasamy:2100}.
In comparing the
results of the scattering calculations to data we found that the grid for the
Fourier transform of the charge symmetry breaking term, which is the
18-th operator in Ref.~\cite{Wiringa:1994wb} needed to be extended beyond 100~fm$^{-1}$,
the standard value for all other terms of the AV18 potential.
Specifically, for the 18th operator  
the Chebyshev expansion of the Fourier transform is extended to 250~fm$^{-1}$.

The scattering amplitude matrix is related to the on-shell
solution of the Lippmann-Schwinger equation as indicated in Eq.~(\ref{g.1}).
Having obtained $M({\bf p}',{\bf p})$,
the spin averaged unpolarized differential cross section is given by 
summing over the final spins and averaging over the initial spins as 
\beq
{d \sigma \over d \Omega} = \frac{1}{4} \mathbf{Tr}(M M^{\dagger}),
\label{g.8}
\eeq
whereas general spin observables have the form  
\beq
\langle O \rangle  = \frac{\mathbf{Tr}(M A M^{\dagger} B)} 
{ \mathbf{Tr}(M M^{\dagger})}.
\label{g.9}
\eeq
Here the operator $A$ is associated with a measurement of the initial
spins and the operator $B$ is associated with a measurement of the final spin
state.  The traces can be expressed as homogeneous polynomials of
degree two in terms of the Wolfenstein parameters. 

To test the potential and the computational method we compare calculated 
$np$ observable as several energies and compare them to the
observables in the GW-DAC analysis~\cite{Arndt:2007qn,GW-DAC} as well as 
calculations based on a 
partial-wave expansion~\cite{Weppner:1997wx}. 
For the $pp$ observables we can only make a comparison with the partial-wave
based calculations, since we do not include the Coulomb interaction. 

\begin{figure}
\begin{center}
\includegraphics[scale=.65]{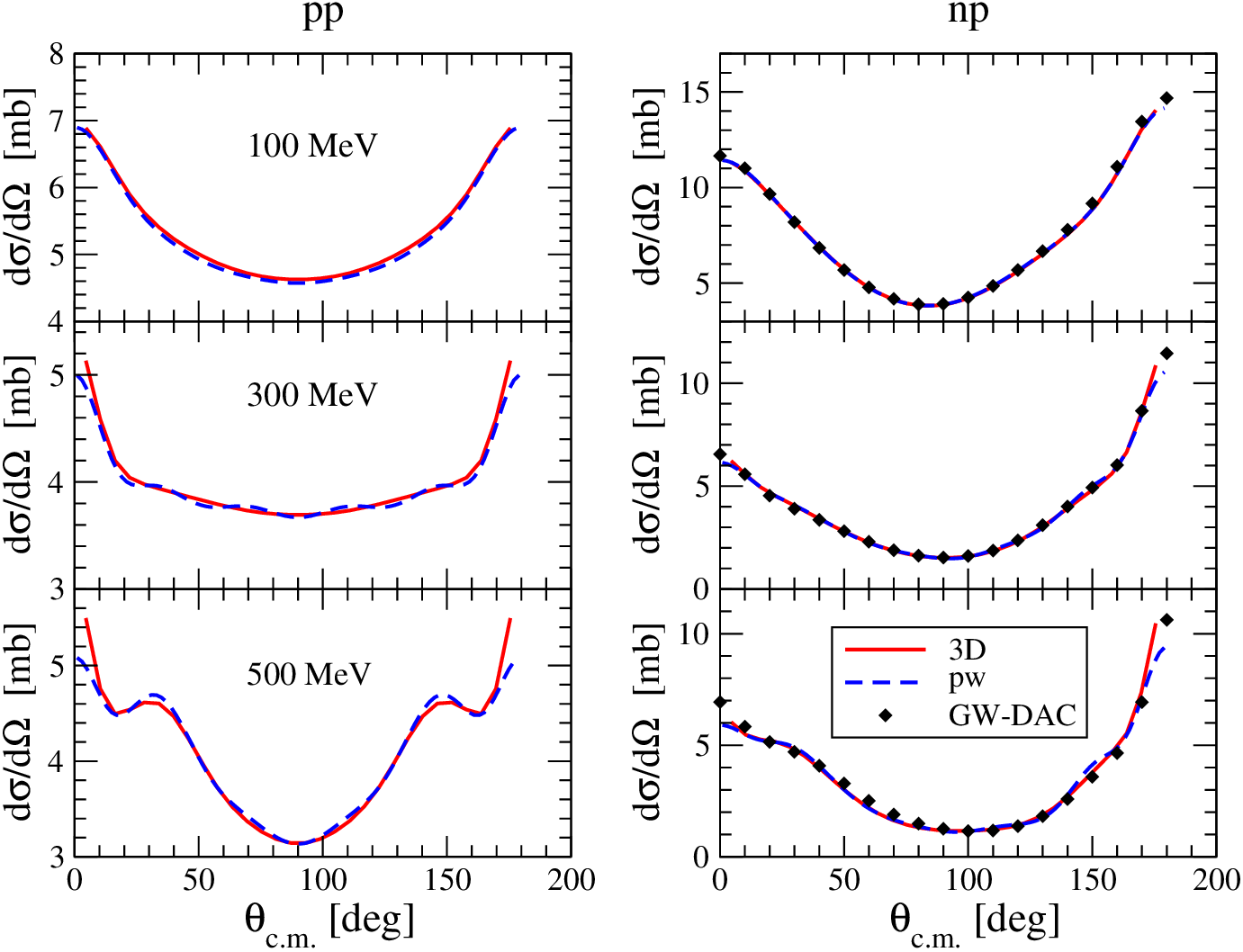}
\caption{The differential cross section for $pp$ (left) and $np$ (right)
scattering as function of the c.m. scattering angle for 100, 300, and 500~MeV
projectile laboratory kinetic energy. The solid (red) line represent the 3D calculation
with the AV18 potential,
whereas the dashed line represent a partial-wave based calculation summed up to
$j_{max}=9$. The diamonds represent the $np$ data from the GW-DAC
analysis~\cite{GW-DAC,Arndt:2007qn}.
}    
\label{fig6}
\end{center}
\end{figure}

Fig.~\ref{fig6} shows the differential cross sections for 
$pp$ and $np$ scattering at 100, 300, and 500~MeV projectile laboratory kinetic
energies.  The solid 
lines are the result of solving the equations of Sec. III with the 
momentum-space AV18 potential of ref. \cite{veerasamy:2100}. 
Both calculations are compared to partial-wave calculations using a partial-wave
sum up to $j_{max}=9$.
The black diamonds are the $np$ cross sections  
taken from the GW-DAC phase shift analysis~\cite{Arndt:2007qn,GW-DAC}.  
For the higher projectile laboratory kinetic energies there is a 
clear deviation at the very forward and backward angles between the 3D
calculation and the partial-wave sum, indicating that even the sum to
$j_{max}=9$ is not yet converged. 

\begin{figure}
\begin{center}
\includegraphics[scale=.65]{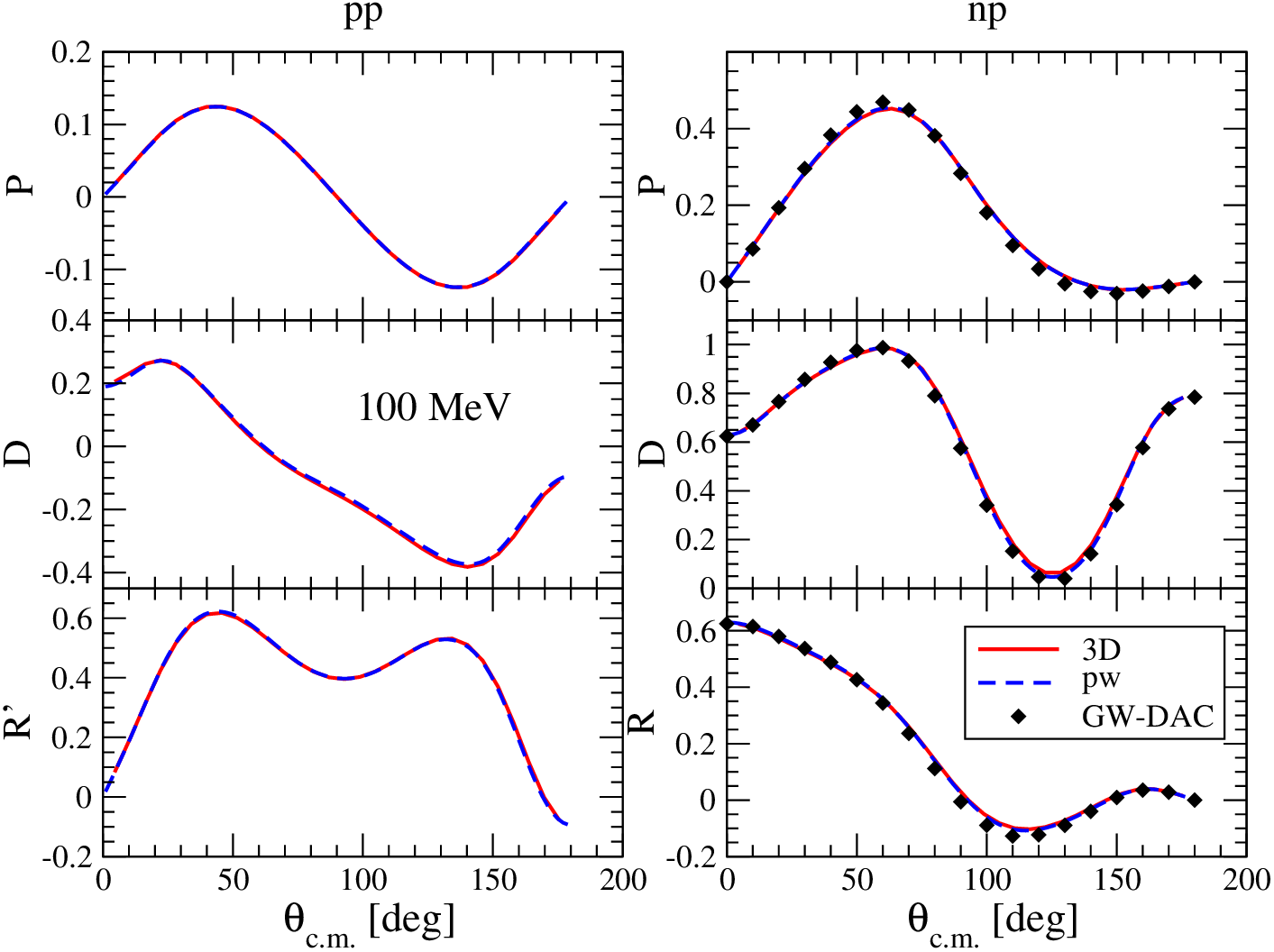}
\caption{The spin observables $P$, $D$ and $R'$ for  $pp$ (left) and $np$ (right)
scattering as function of the c.m. scattering angle for 100~MeV  projectile laboratory
kinetic energy.  The solid (red) line represent the 3D calculation
with the AV18 potential,
whereas the dashed line represent a partial-wave based calculation summed up to
$j_{max}=9$. The diamonds represent the $np$ data from the GW-DAC
analysis~\cite{GW-DAC,Arndt:2007qn}.
}    
\label{fig7}
\end{center}
\end{figure}

\begin{figure}
\begin{center}
\includegraphics[scale=.65]{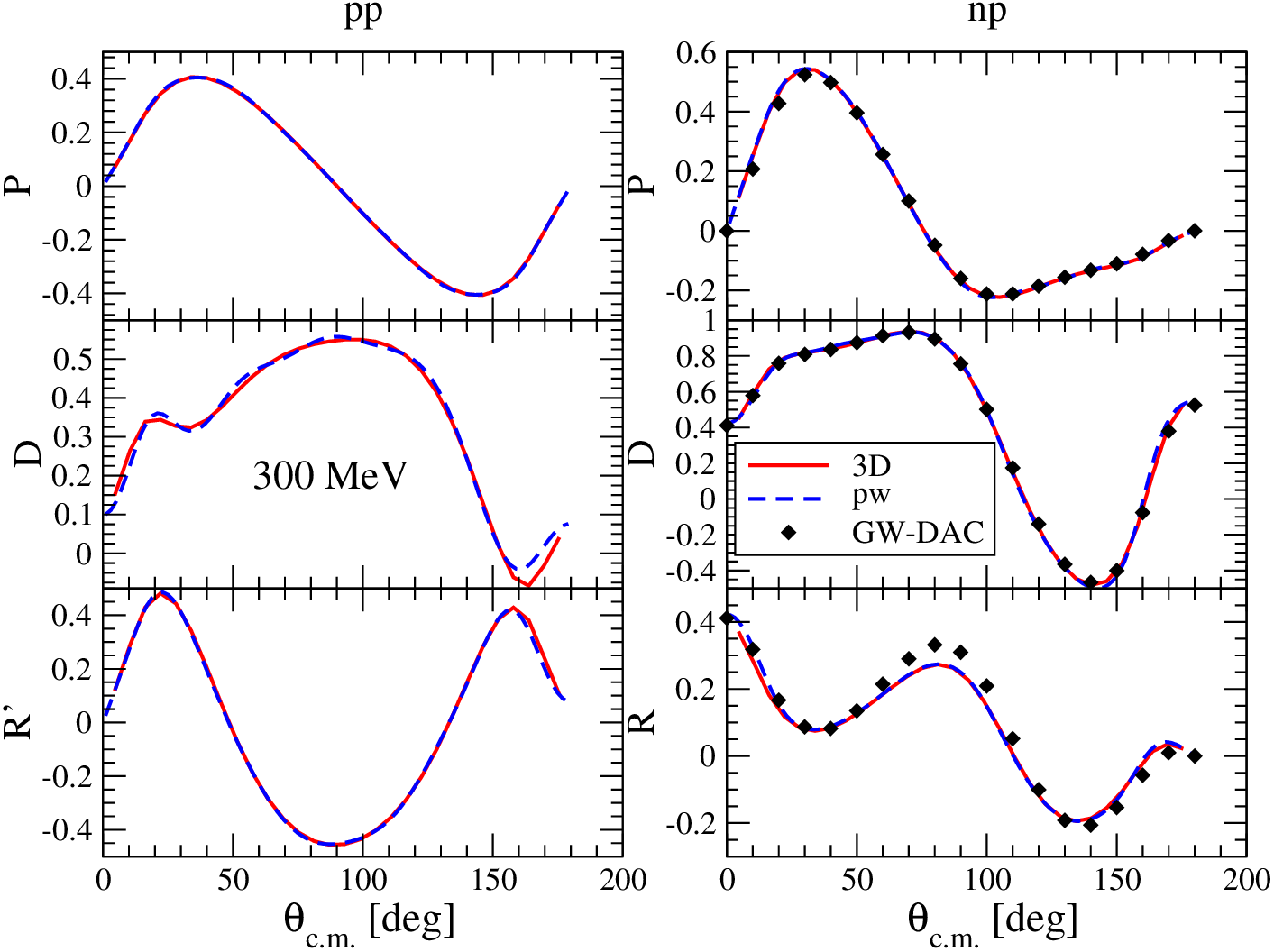}
\caption{The spin observables $P$, $D$ and $R'$ for  $pp$ (left) and $np$ (right)
scattering as function of the c.m. scattering angle for 300~MeV  projectile
laboratory kinetic energy. The meaning of the curves is the same as in
Fig.~\ref{fig7}.
}    
\label{fig8}
\end{center}
\end{figure}

Figs.~\ref{fig7} and \ref{fig8} show selected spin observables for $pp$ and $np$ 
scattering at 
100 and 300~MeV.  The following observables are shown,
\begin{eqnarray}
\mathbf{P} &=& \frac{\mathbf{Tr}(M M^{\dagger} \bsigma)} 
{\mathbf{Tr}(M M^{\dagger})}\\
D &=& \frac{ \mathbf{Tr}(M (\bsigma \cdot \hat{\mathbf{N}}) 
M^{\dagger} (\bsigma \cdot \hat{\mathbf{N}))}} 
{\mathbf{Tr}(M M^{\dagger})} \\
R'& =& \frac{\mathbf{Tr}(M (\bsigma \cdot (\hat{\mathbf{N}} \times \mathbf{Q}))
M^{\dagger} (\bsigma \cdot \hat{\mathbf{Q}}))} 
{ \mathbf{Tr}(M M^{\dagger})}.
\end{eqnarray}
Here $\bsigma$ stands for either $\bsigma_1$ or $\bsigma_2$ depending on
label of the particles.  The solid lines 
represent the 3D calculation with the 
the momentum-space 
AV18 potential of ref. \cite{veerasamy:2100}, whereas the dashed lines 
represent a partial-wave calculation with partial waves summed up to
$j_{max}=9$. The diamonds represent the observables taken from the GW-DAC
analysis~\cite{Arndt:2007qn,GW-DAC}. At 100~MeV projectile kinetic energy there
is perfect agreement between the 3D and the partial-wave calculations, as is
expected. At 300~MeV there are small deviations in the backward angles of $D$
and $R'$, however not as large as for the differential cross section.

\section{Summary and Conclusions}

We formulated and numerically illustrated an approach to treat NN scattering
directly with momentum vectors together with spin-momentum operators multiplied
with scalar function that are only functions of those momentum vectors. Our
formulation differs from the one suggested previously in
Ref.~\cite{Golak:2010wz} in the choice of the operator expansion. The basic
foundation for an operator expansion rests on the fact that the
most general form of a nucleon-nucleon interaction can be represented as a
linear combination of six linearly independent spin-momentum operators with
scalar coefficient functions. The Wolfenstein decomposition of the NN scattering
amplitude into five linearly independent operators is dictated by physical
symmetries~\cite{wolfenstein-ashkin}, while a sixth operator with those symmetry
properties does not exist on-shell. The choice of this sixth operator is not
unique, and different NN potential employ different choices~\cite{Fachruddin:2000wv,Crespo:2001sk}.

In this work we solve the Lippmann Schwinger equation by using
different numbers of spin-momentum basis elements to represent the on- and
half-off shell transition matrix elements.  We show that a numerically
stable discretization of the Lippmann-Schwinger equation is obtained
for a certain pair of on and off-shell spin bases.  The stable choice
includes a sixth time-odd operator in the off-shell basis.
The choice of this sixth operator is the main difference between our
formulation and calculation compared to the ones in
Ref.~\cite{Golak:2010wz}. Our calculations are able to directly
calculate the on-shell elements of the transition amplitude, which was
not easily possible with the choice of Ref.~\cite{Golak:2010wz}. We
should point out that even if a time-odd operator appears in the
expansion of the potential, as it does in the AV18 potential, the
potential itself remains invariant with respect to time reversal,
since the expansion coefficients of this operator also contain
time-odd functions. The scalar coefficient function 
time-odd operator we consider vanishes on
shell.

With those six operators the Lippmann-Schwinger equation becomes a six channel
integral equation in two variables. 
The solution is
represented by six complex amplitudes in four variables, which  on-shell
reduce to five complex amplitudes in two variables, which are simple
linear combinations of the five Wolfenstein amplitudes. We calculated the
Wolfenstein amplitudes for $np$ and $pp$ scattering using the momentum-space
representation of the AV18 potential given in Ref.~\cite{veerasamy:2100} for
several laboratory projectile kinetic energies and compared our calculations to
the results of traditional partial-wave calculations summed up to a given
$j_{max}$ and verified for low energies that our 3D calculation matches the
partial-wave sum exactly, while at even moderately high energies (like e.g.
300~MeV) a partial-wave sum with $j_{max}\leq 9$ is not sufficient for a
converged partial-wave calculation.


\appendix

\section{Transition Matrix Elements}
\label{appendixa}

In order to keep the notation as simple as possible we use  shorthand
notations for the terms in Eqs.~(\ref{e.15}-\ref{e.16}):
\begin{eqnarray}
t_{il}&:=& t(p_i'',p_0,x''_l;p_0)  \cr
t_{0l}&:=& t(p_0,p_0,x'_l;p_0)  \cr
v_{il}&:=& v(p_i',p,x'_l)\cr
v_{0l}&:=& v(p_0 ,p,x'_l) \cr
u_{il;jk}& :=& u(p_i',x'_l,p_j'',x_k'' ) \cr
u_{il;0k}& :=& u(p_i',x'_l,p_0 ,x_k'' )   \cr
u_{0l;jk}& :=& u(p_0,x'_l,p_j'',x_k'' ) \cr
u_{0l;0k} & :=& u(p_0 ,x'_l,p_0 ,x_k'' ) \cr
g_j &:=& {2 \mu \over p_0^2 - p_j^{\prime\prime 2} +i0^+}. 
\end{eqnarray}
In this notation Eqs.~(\ref{e.15}-\ref{e.16}) become
the linear system,
\begin{eqnarray}
t_{il} &=& v_{il}  + \sum_{kj} dx''_k  dp''_j  g_j 
[u_{il;jk}(p''_j)^2 t_{jk} - u_{il;0k}p_0^2 t_{0k}]
- {i 2 \pi \mu p_0 } \sum_k dx''_k u_{il;0k} t_{0k} \cr
t_{0l} &=& v_{0l} +
\sum_{kj} dx''_k  dp''_j  g_j [u_{0l;jk}(p''_j)^2 t_{jk} -u_{0l;0k}p_0^2 t_{0k}] 
- i 2 \pi \mu  p_0   \sum_k dx''_k u_{0l;0k}t_{0k} .
\label{e.27}
\end{eqnarray}
The interaction and transition matrix 
elements are represented by linear combinations of 
the operators, $W^j$.  Since the $W^j$ operators
depend on the momenta, there are different expansions
at each quadrature point.  We use the notation 
$\hat{W}^j$ for the operators used to expand the on-shell 
matrix elements and $W^j$ for the operators used to expand the 
off-shell matrix elements, even though they are identical for
$j=1,\cdots ,5$.   Lower indices are used to denote the momentum 
variables that appear in the operator expressions.
\begin{eqnarray}
t_{il}&:=& \sum_{m=1}^6 t_{il}^m W_{il}^m  \cr
\label{e.28}
t_{0l}&:=& \sum_{m=1}^5 t_{0l}^m \hat{W}_{l}^m \cr
\label{e.29}
v_{il}&:=& \sum_{m=1}^6 v_{il}^m W_{il}^m \cr
\label{e.30}
v_{0l}&:=& \sum_{m=1}^5 v_{0l}^m \hat{W}_{l}^m \cr
\label{e.31}
u_{il;jk}& :=& \sum_{m=1}^6 u_{iljk}^m W_{iljk}^m \cr
\label{e.32}
u_{il;0k}& :=& \sum_{m=1}^6 u_{il0k}^m {W}_{ilk}^m \cr
\label{e.33}
u_{0l;jk}& :=& \sum_{m=1}^6 u_{0l0k}^m {W}_{ilk}^m \cr
\label{e.33b}
u_{0l;0k}& :=& \sum_{m=1}^5 u_{lk}^m \hat{W}_{lk}^m .
\label{e.34}
\end{eqnarray}
Substituting these expansions into Eqs.~(\ref{e.27}) 
leads to
\begin{eqnarray}
\lefteqn{\sum_{m=1}^6  t_{il}^m W_{il}^m  =\sum_{m=1}^6 v_{il}  W_{il}^m + }\cr
& &\sum_{kj} dx_k''  dp_j''  
g_j \left[ \sum_{m,n=1}^6 u^m_{iljk}(p_j^{\prime\prime 2}) t^n_{jk} 
 W_{iljk}^m W_{jk}^n -\sum_{m=1}^6  \sum_{n=1}^5 u^m_{il0k}p_0^2 t^n_{0k}
{W}_{ilk}^m \hat{W}_{k}^n \right] \cr
& &- {i 2 \pi \mu p_0 } \sum_k 
dx_k'' \sum_{m=1}^6\sum_{n=1}^5
u^m_{il0k}t^n_{0k} {W}_{ilk}^m \hat{W}_{k}^n
\label{e.35}
\end{eqnarray}
and
\begin{eqnarray}
\lefteqn{\sum_{m=1}^5 t^m_{0l} \hat{W}_{l}^m =
 \sum_{m=1}^5  v^m _{0l} \hat{W}_{l}^m+ } \cr
& &  \sum_{kj} dx_k''  dp_j''  
g_j \left[\sum_{m,n=1}^6 u^m_{0ljk}(p_j^{\prime\prime 2}) t^n_{jk}
W_{ljk}^m W_{jk}^n - \sum_{m=1}^6\sum_{n=1}^5
u^m_{0l0k}p_0^2 t^n_{0k}  {W}_{lk}^m \hat{W}_{k}^n \right] \cr 
& & - i 2 \pi \mu  p_0   \sum_{k} 
dx_k''\sum_{m,n=1}^5 u^m_{0l0k}t^n_{0k} \hat{W}_{lk}^m \hat{W}_{k}^n
\label{e.36}
\end{eqnarray}
In order to obtain a linear system for the coefficient
functions, $t^m_{il}$ and $t^m_{0l}$, we first  
multiply the first equation from the left by 
$W_{il}^{m'}$ and the second one by  
$\hat{W}_{l}^{m'}$ and then take traces over the spins. 
The resulting linear system is given by  
\begin{eqnarray}
\lefteqn{\sum_{m=1}^6  t_{il}^m \mbox{Tr} (W_{il}^{m'} W_{il}^m)  =
\sum_{m=1}^6 v_{il} \mbox{Tr} (W_{il}^{m'} W_{il}^m) + }\cr
& &\sum_{kj} dx_k''  dp_j''  g_j 
\left[ \sum_{m,n=1}^6 u^m_{iljk}(p_j^{\prime\prime 2}) t^n_{jk} 
\mbox{Tr} (W_{il}^{m'}  W_{iljk}^m W_{jk}^n)
-\sum_{m=1}^6  \sum_{n=1}^5  u^m_{il0k}p_0^2 t^n_{0k}
\mbox{Tr} (W_{il}^{m'} {W}_{ilk}^m \hat{W}_{k}^n) \right] \cr
& & - {i 2 \pi \mu p_0 } \sum_k 
dx_k'' \sum_{m=1}^6\sum_{n=1}^5
u^m_{il0k}t^n_{0k} \mbox{Tr} (W_{il}^{m'} {W}_{ilk}^m \hat{W}_{k}^n)
\label{e.37}
\end{eqnarray}
and
\begin{eqnarray}
\lefteqn{\sum_{m=1}^5 t^m_{0l} \mbox{Tr} (\hat{W}_{l}^{m'} \hat{W}_{l}^m) =
\sum_{m=1}^5  v^m_{0l} \mbox{Tr} (\hat{W}_{l}^{m'} \hat{W}_{l}^m) + } \cr
& &\sum_{kj} dx_k''  dp_j''  g_j \left[\sum_{m,n=1}^6 u^m_{0ljk}
(p_j^{\prime\prime 2}) t^n_{jk}
\mbox{Tr} (\hat{W}_{l}^{m'} W_{ljk}^m W_{jk}^n)
 - \sum_{m=1}^6\sum_{n=1}^5
u^m_{0l0k}p_0^2 t^n_{0k} \mbox{Tr} (\hat{W}_{l}^{m'} {W}_{lk}^m \hat{W}_{k}^n) \right]  \cr
& & - i 2 \pi \mu  p_0   \sum_k 
dx_k''\sum_{m,n=1}^5 u^m_{0l0k}t^n_{0k} \mbox{Tr} (\hat{W}_{l}^{m'} 
\hat{W}_{lk}^m \hat{W}_{k}^n).
\label{e.38}
\end{eqnarray}
The input to Eqs.~(\ref{e.37}) through~(\ref{e.38}) involves a large number of 
traces over the spin operators.  To treat these we use the symbolic 
tool developed in~\cite{saravanan-thesis} to evaluate these traces. 


\begin{acknowledgments}
This work was performed under the auspices of the National Science Foundation
under contract NSF-PHY-1005587 with Ohio University and NSF-PHY-1005501 with the
University of Iowa. Partial support was also provided 
by the U. S. Department of Energy, Office of Nuclear Physics, under contract
No. DE-FG02-93ER40756 with Ohio University, contract
No. DE-FG02-86ER40286 with the University of Iowa.  The authors thank R.B.
Wiringa for providing AV18 phase shifts to cross check calculations of this work.
\end{acknowledgments}


\bibliography{master_bib_file_ch}

\begin{thebibliography}{24}%
\makeatletter
\providecommand \@ifxundefined [1]{%
 \@ifx{#1\undefined}
}%
\providecommand \@ifnum [1]{%
 \ifnum #1\expandafter \@firstoftwo
 \else \expandafter \@secondoftwo
 \fi
}%
\providecommand \@ifx [1]{%
 \ifx #1\expandafter \@firstoftwo
 \else \expandafter \@secondoftwo
 \fi
}%
\providecommand \natexlab [1]{#1}%
\providecommand \enquote  [1]{``#1''}%
\providecommand \bibnamefont  [1]{#1}%
\providecommand \bibfnamefont [1]{#1}%
\providecommand \citenamefont [1]{#1}%
\providecommand \href@noop [0]{\@secondoftwo}%
\providecommand \href [0]{\begingroup \@sanitize@url \@href}%
\providecommand \@href[1]{\@@startlink{#1}\@@href}%
\providecommand \@@href[1]{\endgroup#1\@@endlink}%
\providecommand \@sanitize@url [0]{\catcode `\\12\catcode `\$12\catcode
  `\&12\catcode `\#12\catcode `\^12\catcode `\_12\catcode `\%12\relax}%
\providecommand \@@startlink[1]{}%
\providecommand \@@endlink[0]{}%
\providecommand \url  [0]{\begingroup\@sanitize@url \@url }%
\providecommand \@url [1]{\endgroup\@href {#1}{\urlprefix }}%
\providecommand \urlprefix  [0]{URL }%
\providecommand \Eprint [0]{\href }%
\@ifxundefined \urlstyle {%
  \providecommand \doi  [0]{\begingroup \@sanitize@url \@doi}%
  \providecommand \@doi [1]{\endgroup \@@startlink {\doibase
  #1}doi:\discretionary {}{}{}#1\@@endlink }%
}{%
  \providecommand \doi  [0]{doi:\discretionary{}{}{}\begingroup
  \urlstyle{rm}\Url }%
}%
\providecommand \doibase [0]{http://dx.doi.org/}%
\providecommand \Doi [0]{\begingroup \@sanitize@url \@Doi }%
\providecommand \@Doi  [1]{\endgroup\@@startlink{\doibase#1}\@@Doi}%
\providecommand \@@Doi [1]{#1\@@endlink}%
\providecommand \selectlanguage [0]{\@gobble}%
\providecommand \bibinfo  [0]{\@secondoftwo}%
\providecommand \bibfield  [0]{\@secondoftwo}%
\providecommand \translation [1]{[#1]}%
\providecommand \BibitemOpen [0]{}%
\providecommand \bibitemStop [0]{}%
\providecommand \bibitemNoStop [0]{.\EOS\space}%
\providecommand \EOS [0]{\spacefactor3000\relax}%
\providecommand \BibitemShut  [1]{\csname bibitem#1\endcsname}%
\bibitem [{\citenamefont {Wiringa}\ \emph {et~al.}(1995)\citenamefont
  {Wiringa}, \citenamefont {Stoks},\ and\ \citenamefont
  {Schiavilla}}]{Wiringa:1994wb}%
  \BibitemOpen
  \bibfield  {author} {\bibinfo {author} {\bibfnamefont {R.~B.}\ \bibnamefont
  {Wiringa}}, \bibinfo {author} {\bibfnamefont {V.~G.~J.}\ \bibnamefont
  {Stoks}}, \ and\ \bibinfo {author} {\bibfnamefont {R.}~\bibnamefont
  {Schiavilla}},\ }\Doi {10.1103/PhysRevC.51.38} {\bibfield  {journal}
  {\bibinfo  {journal} {Phys. Rev.},\ }\textbf {\bibinfo {volume} {C51}},\
  \bibinfo {pages} {38} (\bibinfo {year} {1995})},\ \Eprint
  {http://arxiv.org/abs/nucl-th/9408016} {arXiv:nucl-th/9408016} \BibitemShut
  {NoStop}%
\bibitem [{\citenamefont {Machleidt}(2001)}]{Machleidt:2000ge}%
  \BibitemOpen
  \bibfield  {author} {\bibinfo {author} {\bibfnamefont {R.}~\bibnamefont
  {Machleidt}},\ }\Doi {10.1103/PhysRevC.63.024001} {\bibfield  {journal}
  {\bibinfo  {journal} {Phys. Rev.},\ }\textbf {\bibinfo {volume} {C63}},\
  \bibinfo {pages} {024001} (\bibinfo {year} {2001})},\ \Eprint
  {http://arxiv.org/abs/nucl-th/0006014} {arXiv:nucl-th/0006014} \BibitemShut
  {NoStop}%
\bibitem [{\citenamefont {Stoks}\ \emph {et~al.}(1994)\citenamefont {Stoks},
  \citenamefont {Klomp}, \citenamefont {Terheggen},\ and\ \citenamefont
  {de~Swart}}]{Stoks:1994wp}%
  \BibitemOpen
  \bibfield  {author} {\bibinfo {author} {\bibfnamefont {V.}~\bibnamefont
  {Stoks}}, \bibinfo {author} {\bibfnamefont {R.}~\bibnamefont {Klomp}},
  \bibinfo {author} {\bibfnamefont {C.}~\bibnamefont {Terheggen}}, \ and\
  \bibinfo {author} {\bibfnamefont {J.}~\bibnamefont {de~Swart}},\ }\Doi
  {10.1103/PhysRevC.49.2950} {\bibfield  {journal} {\bibinfo  {journal}
  {Phys.Rev.},\ }\textbf {\bibinfo {volume} {C49}},\ \bibinfo {pages} {2950}
  (\bibinfo {year} {1994})},\ \Eprint {http://arxiv.org/abs/nucl-th/9406039}
  {arXiv:nucl-th/9406039 [nucl-th]} \BibitemShut {NoStop}%
\bibitem [{\citenamefont {Epelbaum}\ \emph {et~al.}(2005)\citenamefont
  {Epelbaum}, \citenamefont {Glockle},\ and\ \citenamefont
  {Meissner}}]{Epelbaum:2004fk}%
  \BibitemOpen
  \bibfield  {author} {\bibinfo {author} {\bibfnamefont {E.}~\bibnamefont
  {Epelbaum}}, \bibinfo {author} {\bibfnamefont {W.}~\bibnamefont {Glockle}}, \
  and\ \bibinfo {author} {\bibfnamefont {U.-G.}\ \bibnamefont {Meissner}},\
  }\Doi {10.1016/j.nuclphysa.2004.09.107} {\bibfield  {journal} {\bibinfo
  {journal} {Nucl. Phys.},\ }\textbf {\bibinfo {volume} {A747}},\ \bibinfo
  {pages} {362} (\bibinfo {year} {2005})},\ \Eprint
  {http://arxiv.org/abs/nucl-th/0405048} {arXiv:nucl-th/0405048} \BibitemShut
  {NoStop}%
\bibitem [{\citenamefont {Entem}\ and\ \citenamefont
  {Machleidt}(2003)}]{Entem:2003ft}%
  \BibitemOpen
  \bibfield  {author} {\bibinfo {author} {\bibfnamefont {D.~R.}\ \bibnamefont
  {Entem}}\ and\ \bibinfo {author} {\bibfnamefont {R.}~\bibnamefont
  {Machleidt}},\ }\Doi {10.1103/PhysRevC.68.041001} {\bibfield  {journal}
  {\bibinfo  {journal} {Phys. Rev.},\ }\textbf {\bibinfo {volume} {C68}},\
  \bibinfo {pages} {041001} (\bibinfo {year} {2003})},\ \Eprint
  {http://arxiv.org/abs/nucl-th/0304018} {arXiv:nucl-th/0304018} \BibitemShut
  {NoStop}%
\bibitem [{\citenamefont {Gloeckle}\ \emph {et~al.}(1996)\citenamefont
  {Gloeckle}, \citenamefont {Witala}, \citenamefont {Huber}, \citenamefont
  {Kamada},\ and\ \citenamefont {Golak}}]{Gloeckle:1995jg}%
  \BibitemOpen
  \bibfield  {author} {\bibinfo {author} {\bibfnamefont {W.}~\bibnamefont
  {Gloeckle}}, \bibinfo {author} {\bibfnamefont {H.}~\bibnamefont {Witala}},
  \bibinfo {author} {\bibfnamefont {D.}~\bibnamefont {Huber}}, \bibinfo
  {author} {\bibfnamefont {H.}~\bibnamefont {Kamada}}, \ and\ \bibinfo {author}
  {\bibfnamefont {J.}~\bibnamefont {Golak}},\ }\Doi
  {10.1016/0370-1573(95)00085-2} {\bibfield  {journal} {\bibinfo  {journal}
  {Phys. Rept.},\ }\textbf {\bibinfo {volume} {274}},\ \bibinfo {pages} {107}
  (\bibinfo {year} {1996})}\BibitemShut {NoStop}%
\bibitem [{\citenamefont {Lin}\ \emph {et~al.}(2008){\natexlab{a}}\citenamefont
  {Lin}, \citenamefont {Elster}, \citenamefont {Polyzou}, \citenamefont
  {Witala},\ and\ \citenamefont {Glockle}}]{Lin:2008sy}%
  \BibitemOpen
  \bibfield  {author} {\bibinfo {author} {\bibfnamefont {T.}~\bibnamefont
  {Lin}}, \bibinfo {author} {\bibfnamefont {C.}~\bibnamefont {Elster}},
  \bibinfo {author} {\bibfnamefont {W.~N.}\ \bibnamefont {Polyzou}}, \bibinfo
  {author} {\bibfnamefont {H.}~\bibnamefont {Witala}}, \ and\ \bibinfo {author}
  {\bibfnamefont {W.}~\bibnamefont {Glockle}},\ }\Doi
  {10.1103/PhysRevC.78.024002} {\bibfield  {journal} {\bibinfo  {journal}
  {Phys. Rev.},\ }\textbf {\bibinfo {volume} {C78}},\ \bibinfo {pages} {024002}
  (\bibinfo {year} {2008}{\natexlab{a}})},\ \Eprint
  {http://arxiv.org/abs/0801.3210} {arXiv:0801.3210 [nucl-th]} \BibitemShut
  {NoStop}%
\bibitem [{\citenamefont {Lin}\ \emph {et~al.}(2008){\natexlab{b}}\citenamefont
  {Lin}, \citenamefont {Elster}, \citenamefont {Polyzou},\ and\ \citenamefont
  {Glockle}}]{Lin:2007kg}%
  \BibitemOpen
  \bibfield  {author} {\bibinfo {author} {\bibfnamefont {T.}~\bibnamefont
  {Lin}}, \bibinfo {author} {\bibfnamefont {C.}~\bibnamefont {Elster}},
  \bibinfo {author} {\bibfnamefont {W.~N.}\ \bibnamefont {Polyzou}}, \ and\
  \bibinfo {author} {\bibfnamefont {W.}~\bibnamefont {Glockle}},\ }\Doi
  {10.1016/j.physletb.2008.01.012} {\bibfield  {journal} {\bibinfo  {journal}
  {Phys. Lett.},\ }\textbf {\bibinfo {volume} {B660}},\ \bibinfo {pages} {345}
  (\bibinfo {year} {2008}{\natexlab{b}})},\ \Eprint
  {http://arxiv.org/abs/0710.4056} {arXiv:0710.4056 [nucl-th]} \BibitemShut
  {NoStop}%
\bibitem [{\citenamefont {Veerasamy}\ and\ \citenamefont
  {Polyzou}(2011)}]{veerasamy:2100}%
  \BibitemOpen
  \bibfield  {author} {\bibinfo {author} {\bibfnamefont {S.}~\bibnamefont
  {Veerasamy}}\ and\ \bibinfo {author} {\bibfnamefont {W.~N.}\ \bibnamefont
  {Polyzou}},\ }\href@noop {} {\bibfield  {journal} {\bibinfo  {journal} {Phys.
  Rev.},\ }\textbf {\bibinfo {volume} {C84}},\ \bibinfo {pages} {034003}
  (\bibinfo {year} {2011})}\BibitemShut {NoStop}%
\bibitem [{\citenamefont {Veerasamy}(2011)}]{saravanan-thesis}%
  \BibitemOpen
  \bibfield  {author} {\bibinfo {author} {\bibfnamefont {S.}~\bibnamefont
  {Veerasamy}},\ }\href@noop {} {\bibfield  {journal} {\bibinfo  {journal}
  {Doctoral Dissertation, University of Iowa}} (\bibinfo {year}
  {2011})}\BibitemShut {NoStop}%
\bibitem [{\citenamefont {Fachruddin}\ \emph {et~al.}(2000)\citenamefont
  {Fachruddin}, \citenamefont {Elster},\ and\ \citenamefont
  {Gloeckle}}]{Fachruddin:2000wv}%
  \BibitemOpen
  \bibfield  {author} {\bibinfo {author} {\bibfnamefont {I.}~\bibnamefont
  {Fachruddin}}, \bibinfo {author} {\bibfnamefont {C.}~\bibnamefont {Elster}},
  \ and\ \bibinfo {author} {\bibfnamefont {W.}~\bibnamefont {Gloeckle}},\ }\Doi
  {10.1103/PhysRevC.62.044002} {\bibfield  {journal} {\bibinfo  {journal}
  {Phys. Rev.},\ }\textbf {\bibinfo {volume} {C62}},\ \bibinfo {pages} {044002}
  (\bibinfo {year} {2000})},\ \Eprint {http://arxiv.org/abs/nucl-th/0004057}
  {arXiv:nucl-th/0004057} \BibitemShut {NoStop}%
\bibitem [{\citenamefont {Ramalho}\ \emph {et~al.}(2006)\citenamefont
  {Ramalho}, \citenamefont {Arriaga},\ and\ \citenamefont
  {Pena}}]{Ramalho:2006jk}%
  \BibitemOpen
  \bibfield  {author} {\bibinfo {author} {\bibfnamefont {G.}~\bibnamefont
  {Ramalho}}, \bibinfo {author} {\bibfnamefont {A.}~\bibnamefont {Arriaga}}, \
  and\ \bibinfo {author} {\bibfnamefont {M.~T.}\ \bibnamefont {Pena}},\ }\Doi
  {10.1007/s00601-006-0161-3} {\bibfield  {journal} {\bibinfo  {journal} {Few
  Body Syst.},\ }\textbf {\bibinfo {volume} {39}},\ \bibinfo {pages} {123}
  (\bibinfo {year} {2006})}\BibitemShut {NoStop}%
\bibitem [{\citenamefont {Golak}\ \emph {et~al.}(2010)\citenamefont {Golak},
  \citenamefont {Glockle}, \citenamefont {Skibinski}, \citenamefont {Witala},
  \citenamefont {Rozpedzik} \emph {et~al.}}]{Golak:2010wz}%
  \BibitemOpen
  \bibfield  {author} {\bibinfo {author} {\bibfnamefont {J.}~\bibnamefont
  {Golak}}, \bibinfo {author} {\bibfnamefont {W.}~\bibnamefont {Glockle}},
  \bibinfo {author} {\bibfnamefont {R.}~\bibnamefont {Skibinski}}, \bibinfo
  {author} {\bibfnamefont {H.}~\bibnamefont {Witala}}, \bibinfo {author}
  {\bibfnamefont {D.}~\bibnamefont {Rozpedzik}},  \emph {et~al.},\ }\Doi
  {10.1103/PhysRevC.81.034006} {\bibfield  {journal} {\bibinfo  {journal}
  {Phys.Rev.},\ }\textbf {\bibinfo {volume} {C81}},\ \bibinfo {pages} {034006}
  (\bibinfo {year} {2010})},\ \Eprint {http://arxiv.org/abs/1001.1264}
  {arXiv:1001.1264 [nucl-th]} \BibitemShut {NoStop}%
\bibitem [{\citenamefont {Wolfenstein}\ and\ \citenamefont
  {Ashkin}(1952)}]{wolfenstein-ashkin}%
  \BibitemOpen
  \bibfield  {author} {\bibinfo {author} {\bibfnamefont {L.}~\bibnamefont
  {Wolfenstein}}\ and\ \bibinfo {author} {\bibfnamefont {J.}~\bibnamefont
  {Ashkin}},\ }\href@noop {} {\bibfield  {journal} {\bibinfo  {journal} {Phys.
  Rev.},\ }\textbf {\bibinfo {volume} {85}},\ \bibinfo {pages} {947} (\bibinfo
  {year} {1952})}\BibitemShut {NoStop}%
\bibitem [{\citenamefont {Wolfenstein}(1954)}]{wolfenstein}%
  \BibitemOpen
  \bibfield  {author} {\bibinfo {author} {\bibfnamefont {L.}~\bibnamefont
  {Wolfenstein}},\ }\href@noop {} {\bibfield  {journal} {\bibinfo  {journal}
  {Phys. Rev.},\ }\textbf {\bibinfo {volume} {96}},\ \bibinfo {pages} {1654}
  (\bibinfo {year} {1954})}\BibitemShut {NoStop}%
\bibitem [{\citenamefont {Sloan}(1976)}]{sloan}%
  \BibitemOpen
  \bibfield  {author} {\bibinfo {author} {\bibfnamefont {I.}~\bibnamefont
  {Sloan}},\ }\href@noop {} {\bibfield  {journal} {\bibinfo  {journal} {Math.
  Comp.},\ }\textbf {\bibinfo {volume} {30}},\ \bibinfo {pages} {758} (\bibinfo
  {year} {1976})}\BibitemShut {NoStop}%
\bibitem [{\citenamefont {Keister}\ and\ \citenamefont
  {Polyzou}(2006)}]{Keister:2005eq}%
  \BibitemOpen
  \bibfield  {author} {\bibinfo {author} {\bibfnamefont {B.~D.}\ \bibnamefont
  {Keister}}\ and\ \bibinfo {author} {\bibfnamefont {W.~N.}\ \bibnamefont
  {Polyzou}},\ }\Doi {10.1103/PhysRevC.73.014005} {\bibfield  {journal}
  {\bibinfo  {journal} {Phys. Rev.},\ }\textbf {\bibinfo {volume} {C73}},\
  \bibinfo {pages} {014005} (\bibinfo {year} {2006})},\ \Eprint
  {http://arxiv.org/abs/nucl-th/0508001} {arXiv:nucl-th/0508001} \BibitemShut
  {NoStop}%
\bibitem [{\citenamefont {Reed}\ and\ \citenamefont {Simon}(1972)}]{simon1}%
  \BibitemOpen
  \bibfield  {author} {\bibinfo {author} {\bibfnamefont {M.}~\bibnamefont
  {Reed}}\ and\ \bibinfo {author} {\bibfnamefont {B.}~\bibnamefont {Simon}},\
  }\href@noop {} {\emph {\bibinfo {title} {Methods of Modern mathematical
  Physics}}},\ Vol.\ \bibinfo {volume} {I Functional Analysis}\ (\bibinfo
  {publisher} {Academic Press},\ \bibinfo {year} {1972})\BibitemShut {NoStop}%
\bibitem [{\citenamefont {Gloeckle}(1983)}]{glockle:1983}%
  \BibitemOpen
  \bibfield  {author} {\bibinfo {author} {\bibfnamefont {W.}~\bibnamefont
  {Gloeckle}},\ }\href@noop {} {\emph {\bibinfo {title} {The Quantum Mechanical
  Few-Body Problem}}}\ (\bibinfo  {publisher} {Springer-Verlag},\ \bibinfo
  {year} {1983})\BibitemShut {NoStop}%
\bibitem [{\citenamefont {Weppner}\ \emph {et~al.}(1998)\citenamefont
  {Weppner}, \citenamefont {Elster},\ and\ \citenamefont
  {Huber}}]{Weppner:1997wx}%
  \BibitemOpen
  \bibfield  {author} {\bibinfo {author} {\bibfnamefont {S.}~\bibnamefont
  {Weppner}}, \bibinfo {author} {\bibfnamefont {C.}~\bibnamefont {Elster}}, \
  and\ \bibinfo {author} {\bibfnamefont {D.}~\bibnamefont {Huber}},\ }\Doi
  {10.1103/PhysRevC.57.1378} {\bibfield  {journal} {\bibinfo  {journal}
  {Phys.Rev.},\ }\textbf {\bibinfo {volume} {C57}},\ \bibinfo {pages} {1378}
  (\bibinfo {year} {1998})},\ \Eprint {http://arxiv.org/abs/nucl-th/9712001}
  {arXiv:nucl-th/9712001 [nucl-th]} \BibitemShut {NoStop}%
\bibitem [{\citenamefont {Arndt}\ \emph {et~al.}(2007)\citenamefont {Arndt},
  \citenamefont {Briscoe}, \citenamefont {Strakovsky},\ and\ \citenamefont
  {Workman}}]{Arndt:2007qn}%
  \BibitemOpen
  \bibfield  {author} {\bibinfo {author} {\bibfnamefont {R.}~\bibnamefont
  {Arndt}}, \bibinfo {author} {\bibfnamefont {W.}~\bibnamefont {Briscoe}},
  \bibinfo {author} {\bibfnamefont {I.}~\bibnamefont {Strakovsky}}, \ and\
  \bibinfo {author} {\bibfnamefont {R.}~\bibnamefont {Workman}},\ }\href@noop
  {} {\bibfield  {journal} {\bibinfo  {journal} {Phys.Rev.},\ }\textbf
  {\bibinfo {volume} {C76}},\ \bibinfo {pages} {025209} (\bibinfo {year}
  {2007})},\ \Eprint {http://arxiv.org/abs/0706.2195} {arXiv:0706.2195
  [nucl-th]} \BibitemShut {NoStop}%
\bibitem [{\citenamefont {Arndt}\ \emph {et~al.}(2009)\citenamefont {Arndt},
  \citenamefont {Briscoe}, \citenamefont {Laptev}, \citenamefont {Strakovsky},\
  and\ \citenamefont {Workman}}]{Arndt:2008uc}%
  \BibitemOpen
  \bibfield  {author} {\bibinfo {author} {\bibfnamefont {R.}~\bibnamefont
  {Arndt}}, \bibinfo {author} {\bibfnamefont {W.}~\bibnamefont {Briscoe}},
  \bibinfo {author} {\bibfnamefont {A.}~\bibnamefont {Laptev}}, \bibinfo
  {author} {\bibfnamefont {I.}~\bibnamefont {Strakovsky}}, \ and\ \bibinfo
  {author} {\bibfnamefont {R.}~\bibnamefont {Workman}},\ }\href@noop {}
  {\bibfield  {journal} {\bibinfo  {journal} {Nucl.Sci.Eng.},\ }\textbf
  {\bibinfo {volume} {162}},\ \bibinfo {pages} {312} (\bibinfo {year}
  {2009})},\ \Eprint {http://arxiv.org/abs/0806.1198} {arXiv:0806.1198
  [nucl-ex]} \BibitemShut {NoStop}%
\bibitem [{\citenamefont {http://gwdac.phys.gwu.edu}()}]{GW-DAC}%
  \BibitemOpen
  \bibfield  {author} {\bibinfo {author} {\bibnamefont
  {http://gwdac.phys.gwu.edu}}\ }
\bibitem [{\citenamefont {Crespo}\ and\ \citenamefont
  {Moro}(2002)}]{Crespo:2001sk}%
  \BibitemOpen
  \bibfield  {author} {\bibinfo {author} {\bibfnamefont {R.}~\bibnamefont
  {Crespo}}\ and\ \bibinfo {author} {\bibfnamefont {A.~M.}\ \bibnamefont
  {Moro}},\ }\Doi {10.1103/PhysRevC.65.054001} {\bibfield  {journal} {\bibinfo
  {journal} {Phys. Rev.},\ }\textbf {\bibinfo {volume} {C65}},\ \bibinfo
  {pages} {054001} (\bibinfo {year} {2002})},\ \Eprint
  {http://arxiv.org/abs/nucl-th/0110003} {arXiv:nucl-th/0110003} \BibitemShut
  {NoStop}%
\end{thebibliography}%

\clearpage

\end{document}